\newcommand{\lya}{\mbox{Ly$\alpha$}}

\documentclass[twocolumn]{aastex631}

\usepackage{graphicx}

\received{Int' nu}
\accepted{XXX}
\submitjournal{ApJ}

\shorttitle{An unbiased \lya-emitter survey at $z>7.5$}
\shortauthors{A. Runnholm et al.}

\begin{document}

\title{The JWST/PASSAGE Survey: Testing Reionization Histories with JWST’s First Unbiased Survey for Lyman alpha Emitters at Redshifts 7.5--9.5}

\author[0000-0002-1025-7569]{Axel Runnholm}
\affiliation{Stockholm University, Department of Astronomy and Oskar Klein Centre for Cosmoparticle Physics, AlbaNova University Centre, SE-10691, Stockholm, Sweden.}

\author[0000-0001-8587-218X]{Matthew J. Hayes}
\affiliation{Stockholm University, Department of Astronomy and Oskar Klein Centre for Cosmoparticle Physics, AlbaNova University Centre, SE-10691, Stockholm, Sweden.}


\author[0000-0001-7166-6035]{Vihang Mehta}
\affiliation{IPAC, Mail Code 314-6, California Institute of Technology, 1200 E. California Blvd., Pasadena, CA, 91125, USA}


\author[0000-0001-6919-1237]{Matthew A. Malkan}
\affiliation{University of California, Los Angeles, Department of Physics and Astronomy, 430 Portola Plaza, Los Angeles, CA 90095, USA}


\author[0000-0002-9136-8876]{Claudia Scarlata}
\affiliation{Minnesota Institute for Astrophysics, University of Minnesota, 116 Church Street SE, Minneapolis, MN 55455, USA}

\author[0000-0001-5294-8002]{Kalina~V.~Nedkova} 
\affiliation{Department of Physics and Astronomy, Johns Hopkins University, Baltimore, MD 21218, USA}
\affiliation{Space Telescope Science Institute, 
3700 San Martin Drive, 
Baltimore, MD, 21218 USA}

\author[0000-0002-9946-4731]{Marc Rafelski}
\affiliation{Space Telescope Science Institute, 
3700 San Martin Drive, 
Baltimore, MD, 21218 USA}
\affiliation{Department of Physics and Astronomy, Johns Hopkins University, Baltimore, MD 21218, USA}

\author[0000-0003-0980-1499]{Benedetta Vulcani}
\affiliation{INAF, Osservatorio Astronomico di Padova, Vicolo dell'Osservatorio 5, 35122 Padova, Italy}

\author[0009-0002-9932-4461]{Mason Huberty}
\affiliation{Minnesota Institute for Astrophysics, University of Minnesota, 116 Church Street SE, Minneapolis, MN 55455, USA}

\author{E. Christian Herenz}
\affiliation{Inter-University Centre for Astronomy and Astrophysics (IUCAA), Pune University Campus, Pune 411 007, India}

\author{Anne Hutter }
\affiliation{Niels Bohr Institute, University of Copenhagen, Jagtvej 128, DK-2200, Copenhagen N, Denmark}
\affiliation{Cosmic Dawn Center (DAWN)}

\author[0000-0002-6503-5218]{Sean Bruton}
\affiliation{California Institute of Technology, 1200 E. California Blvd., Pasadena, CA, 91125, USA}


\author[0000-0003-4804-7142]{Ayan Acharyya}
\affiliation{INAF, Osservatorio Astronomico di Padova, Vicolo dell'Osservatorio 5, 35122 Padova, Italy}

\author[0000-0002-7570-0824]{Hakim Atek}
\affiliation{Institut d'Astrophysique de Paris, CNRS, Sorbonne Universit\'e, 98bis Boulevard Arago, 75014, Paris, France}

\author[0000-0003-4804-7142]{Ivano Baronchelli}
\affiliation{INAF -Istituto di Radioastronomia, Via Gobetti 101 40129 Bologna, Italy}

\author[0000-0003-4569-2285]{Andrew J. Battisti}
\affiliation{International Centre for Radio Astronomy Research (ICRAR), University of Western Australia, M468, 35 Stirling Highway, Crawley, WA 6009, Australia}
\affiliation{Australian National University, Research School of Astronomy and Astrophysics, Canberra, ACT 2611, Australia}
\affiliation{ARC Centre of Excellence for All Sky Astrophysics in 3 Dimensions (ASTRO 3D), Australia}

\author[0000-0001-5984-0395]{Maru\v{s}a Brada{\v c}}
\affiliation{University of Ljubljana, Department of Mathematics and Physics, Jadranska ulica 19, SI-1000 Ljubljana, Slovenia}
\affiliation{Department of Physics and Astronomy, University of California Davis, 1 Shields Avenue, Davis, CA 95616, USA}

\author[0000-0002-8651-9879]{Andrew J.\ Bunker}
\affiliation{Department of Physics, University of Oxford, Denys Wilkinson Building, Keble Road, Oxford OX1 3RH, UK}

\author[0000-0002-7928-416X]{Y. Sophia Dai}
\affiliation{The National Astronomical Observatories, Chinese Academy of Sciences, 20A Datun Road, Chaoyang District, Beijing 100101, China}

\author[0009-0009-0868-8165]{Clea Hannahs}
\affiliation{University of California, Los Angeles, Department of Physics and Astronomy, 430 Portola Plaza, Los Angeles, CA 90095, USA}

\author[0000-0002-0072-0281]{Farhanul Hasan}
\affiliation{Space Telescope Science Institute, 
3700 San Martin Drive, 
Baltimore, MD, 21218 USA}

\author[0000-0001-6505-0293]{Keunho J. Kim}
\affiliation{IPAC, Mail Code 314-6, California Institute of Technology, 1200 E. California Blvd., Pasadena, CA, 91125, USA}

\author[0000-0003-4570-3159]{Nicha Leethochawalit}
\affiliation{National Astronomical Research Institute of Thailand (NARIT), Mae Rim, Chiang Mai, 50180, Thailand}

\author[0000-0001-8792-3091]{Yu-Heng Lin}
\affiliation{IPAC, Mail Code 314-6, California Institute of Technology, 1200 E. California Blvd., Pasadena, CA, 91125, USA}

\author[0000-0001-7016-5220]{Michael J. Rutkowski}
\affiliation{Department of Physics and Astronomy, Minnesota State University, Mankato, Trafton North 158, Mankato, MN 56001, USA}

\author[0000-0001-8419-3062]{Alberto Saldana-Lopez}
\affiliation{Stockholm University, Department of Astronomy and Oskar Klein Centre for Cosmoparticle Physics, AlbaNova University Centre, SE-10691, Stockholm, Sweden.}

\author[0000-0002-0364-1159]{Zahra Sattari}
\affiliation{IPAC, Mail Code 314-6, California Institute of Technology, 1200 E. California Blvd., Pasadena, CA, 91125, USA}

\author[0000-0002-9373-3865]{Xin Wang}
\affiliation{School of Astronomy and Space Science, University of Chinese Academy of Sciences (UCAS), Beijing 100049, China}
\affiliation{National Astronomical Observatories, Chinese Academy of Sciences, Beijing 100101, China}
\affiliation{Institute for Frontiers in Astronomy and Astrophysics, Beijing Normal University, Beijing 102206, China}

\begin{abstract}
Lyman $\alpha$ (Ly$\alpha$) emission is one of  few observable features of galaxies that can trace the neutral hydrogen content in the Universe during the Epoch of Reionization (EoR). To accomplish this we need an efficient way to survey for Ly$\alpha$ emitters (LAEs) at redshifts beyond 7, requiring unbiased emission-line observations that are both sufficiently deep and wide to cover enough volume to detect them. Here we present results from PASSAGE---a pure-parallel JWST/NIRISS slitless spectroscopic survey to detect Ly$\alpha$ emitters deep into the EoR, without the bias of photometric preselection. We identify four LAEs at $7.5\leq z\leq9.5$ in four surveyed pointings, and  estimate the luminosity function (LF).  We find that the LF does show a marked decrease compared to post-reionization measurements, but the change is a factor of $\lesssim 10$, which is less than expected from theoretical calculations and simulations, as well as  observational expectations from the pre-JWST literature. Modeling of the IGM and expected \lya\ profiles implies these galaxies reside in ionized bubbles of $\gtrapprox 2$ physical Mpc.  We also report that in the four fields we detect \{3,1,0,0\} LAEs, which could indicate strong field-to-field variation in the LAE distribution, consistent with a patchy HI distribution at $z\sim8$.  We compare the recovered LAE number counts with expectations from simulations and discuss the potential implications  for reionization and its morphology. 
\end{abstract}

\keywords{galaxies: high-redshift; galaxies: Lyman-alpha emitters; intergalactic medium; reionization}

\section{Introduction} \label{sec:intro}
The Epoch of Reionization (EoR) is a time of great astrophysical interest since it represents both a major phase transition of the universe as well as the earliest stages of galaxy evolution \citep[see e.g.][for a review]{robertson2022a}. Despite this, relatively little is certain about how the transition proceeded, especially at redshifts ($z$) beyond 7. The Planck \citep{planckcollaboration2020} experiment used electron scattering measurements to constrain the midpoint of the EoR to be $z \sim 7.5$. Observations of the Lyman $\alpha$ (Ly$\alpha$) forest and damping wing observed in quasar spectra have placed the end of reionization between redshift 5 and 6 \citep{kulkarni2019,keating2020,bosman2022, becker2024, qin_mesinger2024, zhu2024, grieg2024}. However, the start of the EoR, and whether it proceeded rapidly or more gradually, is still debated \citep{naidu2020, finkelstein2019, bolan2022, napolitano2024a, becker2024} . In fact the shape of the transition itself may have strong implications for the sources that powered it, both in terms of power, number density and spatial distribution and even for their nature \citep{qin2024}.

The neutral intergalactic medium (IGM) consists mostly of atomic hydrogen, which makes studying it directly extremely challenging. The Square Kilometer Array  \citep[SKA;][]{Dewdney09, koopmans2015} will attempt to target 21cm emission from this epoch, however, even with this new facility, the problem will remain a difficult one. Therefore, we must turn to additional, more indirect, tracers to better constrain the reionization history.  One such tracer is Ly$\alpha$ emission, which is an intrinsically strong emission line ($\sim68$\% of ionizing photons result in Ly$\alpha$ emission under normal gas conditions) from recombining hydrogen gas \citep[see e.g.][]{dijkstra2014,ouchi2020}. As a result of its resonant nature Ly$\alpha$ radiation interacts strongly with neutral atomic hydrogen gas. Therefore, in principle, it can be used to trace the column of atomic gas in the IGM and hence neutral fraction of the Universe. 

This has been done using various metrics, including the evolution of the Ly$\alpha$ luminosity function (LF, \citealp{kashikawa2011, wold2022}), the volume-averaged escape fraction \citep{hayes2011}, the Ly$\alpha$ emitter (LAE) fraction \citep{stark2010, stark2011, pentericci2011, ono2012, kusakabe2022}, and the equivalent width distribution \citep{mason2018, mason2019, bolan2022}. Previous works, such as \citet{ouchi2008, herenz2019}, have shown that the Ly$\alpha$ LF remains relatively constant between redshift 3 to 6, although some changes at the extreme ends of the LF have been noted \citep[e.g.][]{thai2023} . However, observing Ly$\alpha$ emitters and, consequently, LFs at higher redshifts has proven challenging, but not impossible using ground-based narrowband searches. Early attempts indicated a dearth of Ly$\alpha$ detections \citep{cuby2007, willis2008}, which on its own suggested a significant drop in the LAE density at redshifts above 6. More recently, \citet{itoh2018},  and\citet{wold2022}, detected small numbers of LAEs, and confirm the reduction of the Ly$\alpha$ LF at $z\sim7$ . A significant reduction in the Ly$\alpha$ LF at the same $z$ was also found by the SILVERRUSH collaboration \citep{ouchi2018, kikuta2023, umeda2024b}.

Nevertheless, a reduction of the Ly$\alpha$ LF is insufficient to conclude that the Ly$\alpha$ optical depth is increasing purely because of a neutral IGM. This could also be due to the galaxy population being different at higher redshifts. Therefore, IGM inferences have relied on various comparisons to the UV properties of the galaxy population at the same redshift. For example, \citet{hayes2011}, \citet{itoh2018} and \citet{wold2022} compare the implied Ly$\alpha$ luminosity density (LD) evolution with redshift to the UV LD evolution. \citet{itoh2018} found that Ly$\alpha$ evolves faster---an indication of neutral IGM damping wing absorption, whereas \citet{wold2022} conclude that there is no indication of this at $z\sim7$, implying a fully ionized IGM. \citet{mason2018, mason2019} and \citet{bolan2022}, on the other hand, rely on modeling the Ly$\alpha$ equivalent width (EW) distribution relation to the UV LF and conclude that the neutral fraction, $x_{HI}$, is about 0.6 at redshift 7, whereas the results of \citet{mason2025} indicate that the universe reaches those $x_{HI}$ closer to $z\sim9$. Clearly, there are large uncertainties and discrepancies in EoR progression inferences and more observations are needed, especially at higher redshift to put the conclusions on solid footing.

The high sensitivity infrared capabilities of JWST have opened a new window into this and are enabling the detection of Ly$\alpha$ emitters at unprecedented redshifts. For instance \citet{nakane2024} collected 53 galaxies with confirmed Ly$\alpha$ emission from GLASS \citep{Treu2022ApJ}, CEERS \citep{finkelstein2023a}, and JADES \citep{Bunker2024A&A, Eisenstein2023arXiv} above $z\geq7$, and \citet{tang2024a} presents a compilation of more than 210 LAEs at $z\geq6.5$. Such datasets have already been used to estimate the sizes of ionized regions around the observed galaxies, finding necessary radii of 0.1-1\,pMpc \citep{ witstok2024a, witstok2025} as well as tentative growth of ionized regions with decreasing redshift \citep[][]{hayesscarlata2023}. \citet{witstok2024b} recently reported the detection of a strong Ly$\alpha$ line at an impressively high redshift of 13, which was wholly unexpected due to the high IGM neutral fractions expected to efficiently scatter Ly$\alpha$ at these redshifts.

Using such observations we are now able to start pushing EoR constraints from the end stages all the way to when the process began, and determine the reionization history. Different sources of reionization have different implications for how the process proceeds, whether it is powered by a few bright galaxies (oligarchal reionization, \citealt{naidu2020, sharma2016}) or by emission from many faint galaxies  (also referred to as democratic reionization, \citealp{finkelstein2019,atek24}) will affect how gradual it is \citep{qin2024}, as well as its topology. Measuring this well will require large unbiased samples of LAEs in the EoR---in this work we demonstrate the feasibility of obtaining such samples with JWST/NIRISS.

One limitation of current high-redshift Ly$\alpha$ samples is that they are derived from a small number of relatively small-area surveys. This means that any inferences from them are potentially susceptible to cosmic variance, and may be part of the explanation of why Ly$\alpha$-based $x_{HI}$ determinations at the same redshifts may disagree with each other\citep[see e.g.][]{napolitano2024a}. The most effective way to overcome cosmic variance is to observe many uncorrelated fields rather than expanding a single contiguous field \citep{bruton2023a}. This is the primary driver of the Cycle 1 JWST program PASSAGE (Parallel Application of Slitless Spectroscopy for the Analysis of Galaxy Evolution, Malkan et al.~submitted.) which uses pure parallel observations with NIRISS wide field slitless spectroscopy (WFSS) to conduct an unbiased search for emission line galaxies. 
In this work we develop a technique for pure emission line selection in grism spectroscopy, and use the deepest PASSAGE data to search for faint Ly$\alpha$ emission to build up a Ly$\alpha$ LF at $7.5\leq z \leq 9.5$. 

In Section\,\ref{sec: grism motivation} we motivate why WFSS is the most efficient way to perform an unbiased Ly$\alpha$ survey at $z>7$. In Section\,\ref{sec:data} we describe the observations and data reduction, as well as the bespoke process of source detection, calibration, and validation.  We describe the selected LAE candidates in Section\,\ref{sec:sample}, discuss the implications of our observations for reionization in Section\,\ref{sec:reionization}. Sections\,\ref{sec:discussion} and \ref{sec:conclusions} give a broader discussion and conclusions. We use a concordance cosmology ($\Omega_M =0.3$, $\Omega_\Lambda=0.7$, $H_0=70$) throughout. 

\section{Grism-based emission line selection}\label{sec: grism motivation}

Although \lya-emitters are now being identified at $z>7$, charaterizing the population as a whole is not straightforward. The reason for this is that all nebular \lya\ from these redshifts is identified by  follow-up spectroscopy of continuum-selected galaxies. This introduces a bias, as the resulting LF is sampled as the UV LF convolved with the equivalent width distribution \citep[see e.g.][for discussion]{dijkstra2012, gronke2015} rather than being derived from Ly$\alpha$ emission directly.  This makes it difficult, if not impossible, to quantify the sample completeness in Ly$\alpha$ luminosities. 

Ideally, we should select LAEs directly  by their Ly$\alpha$ emission. Many Ly$\alpha$ LF studies have done this either using IFU spectroscopy below redshift $\sim6.6$ \citep[e.g.][]{herenz2019}, or at higher redshifts, using narrowband selection from the ground, \citep[see e.g.][]{cuby2007,willis2008}. At $z\gtrsim 6.5$ these studies become very challenging and the reachable depths are heavily limited by the sky background---skylines, absorption, and IR background impact sensitivity as well as available $z$-range which limits the observable volume. Though some campaigns, for instance UltraVISTA \citep{mccracken2012, laursen2019}, have targeted large sky areas to mitigate this, line sensitivity remains low. Furthermore, low-$z$ interlopers, such as [OII], [OIII] and H$\alpha$-emitters limit the purity of the samples. At Earth--Sun L2, JWST NIR observations do not suffer from airglow background, which allows much greater depths to be reached in comparable short integration times. Additionally, the use of NIRISS' wide field slitless spectroscopic (WFSS) capabilities lets us cover a much larger range in redshift in a single observation than a narrowband filter, and provides better interloper rejection, due to the longer wavelength coverage in which other emission lines can be identified. Together they enable us to conduct efficient, unbiased near infrared (NIR) emission line surveys for the first time.

In this work we set out to derive the $7.5\leq z \leq 9.5$ Ly$\alpha$ LF from NIRISS WFSS data. However, standard WFSS processing uses a direct detection image to do pre-selection before spectral extraction. This means that in many cases `emission line galaxies' are actually continuum- rather than line-selected, and the samples suffer the same bias as discussed above. This becomes an even more pronounced issue when using parallel observations, where the direct image covering the emission line may be shallow or even missing. Therefore we develop methods in this paper to selecting emission line galaxies directly in the dispersed images, adopting the philosophy of \citet{atek2010} \citep[see also][]{bagley2017, battisti2024}. We describe the procedure in detail in Section\,\ref{sec: spectrum extraction}. This allows us to do a pure emission line selected Ly$\alpha$ survey at high-$z$. 

\section{Observations and Data Processing} \label{sec:data}

Here we describe the technical details of the observations, how we extract spectra, and verify our LAE candidates. Very briefly, we do this by searching for emission lines in the GR150C grism data using the F115W filter, and confirming that the same line exists in the GR150R grism direction.  We then verify that there is no detection in deep ancillary imaging data blueward of the emission line, and that no other emission lines are present in the rest of the spectral region covered by the grism data, including observations in the F150W and F200W filters. We specifically search for single, symmetric, emission lines since the characteristic Ly$\alpha$ spectral profile is unresolved by NIRISS and the [OIII]$_{4959, 5007}$ line commonly shows an asymmetric profile. Readers interested primarily in the results may skip to Section\,\ref{sec:sample}.

\begin{deluxetable*}{lccc|ccc|ccc|ccc}\label{tab:observations}
\tabletypesize{\scriptsize}
\tablewidth{0pt} 
\tablecaption{Observational log}
\tablehead{
\colhead{Parallel} & \multicolumn{3}{c}{F115W}& \multicolumn{3}{c}{F150W} & \multicolumn{3}{c}{F200W} & \multicolumn{3}{c}{Ancillary data}\\
\colhead{} & \colhead{Direct} & \colhead{GR150C} & \colhead{GR150R}  & \colhead{Direct} & \colhead{GR150C} & \colhead{GR150R}  & \colhead{Direct} & \colhead{GR150C} & \colhead{GR150R}
& \colhead{Instrument} & \colhead{Filter} & \colhead{1$\sigma$ depth } }
\startdata 
Par 28 & 6442 & 34014 & 42517 & 5153 & 8503  & 17007 & 5153 & 6442  & 12884 & HST/ACS & F814W & 29.5\\  
Par 50 & --   & 21645 & 21645 & 7344 & --    & 14173 & 7344 & --    & 14172 & Suprime-Cam$^1$ & R-band& 28.5\\  
Par 52 & --   & 21645 & 21645 & 7344 & --    & 14173 & 7344 & --    & 14172 & Suprime-Cam$^1$ & R-band& 30.3\\    
NGDEEP & 5411 & 46962 & 46962 & 1803 & 20872 & 20872 & 1803 & 15654 & 15654 & HST/ACS & F814W& 32.1\\  
\enddata
\tablecomments{Integration times are given in seconds, and limiting magnitudes in the AB system \citep{oke1983}. $^1$ Suprime-Cam data was retrieved from the Suprime-Cam Legacy Archive \citep{gwyn2020} (\url{https://www.cadc-ccda.hia-iha.nrc-cnrc.gc.ca/en/scla/}). }
\end{deluxetable*}

\subsection{Observations\label{subsec:observations}}
The main dataset for this study comes from the PASSAGE survey (Malkan et al. submitted), which consists in total of 63 parallel pointings observed with NIRISS with various combinations of the F115W, F150W and F200W filters and a wide range of depths. Here we search for high-redshift Ly$\alpha$ emitting sources, which are naturally faint both in continuum and the Ly$\alpha$ line. This places a limit on which fields can be used, since many observations are quite shallow with less than 1.5 hours of exposure time per grism and filter combination. For this survey we require a minimum of 2 hours of exposure time in each dispersion direction in the F115W filter. Since we are working at close to the detection limit of the data, we require coverage in both NIRISS grism orients---GR150C and GR150R--- to confirm the line detections and establish a wavelength calibration. Together, these requirements leave four initial parallel fields that are of interest for this work. In addition to the PASSAGE data we also use archival NIRISS data in the same WFSS configurations from the NGDEEP program \citep{bagley2024}, which covers the Hubble Ultra Deep Field. 

In order to further validate single line emitters as LAEs we also need deep ancillary data that lies blueward of our F115W data, in order to verify that there is no emission blueward of the emission line---as expected for Ly$\alpha$ emitters (see Section\,\ref{sec: validation} for details). We search for such imaging in both HST archives and the Suprime-Cam Legacy Archive. For one field only very shallow SDSS imaging exist, which means we cannot exclude low-$z$ interlopers, and therefore we do not consider this field further. This leaves us with a total of three PASSAGE fields---Par 28, 50, and 52 in the nomenclature of the APT file---and the NGDEEP field covering a total area of $\approx 19$ square arcmin. A summary of the exposure times of the four fields is given in Table~\ref{tab:observations}. The last two columns of Table~\ref{tab:observations} detail the ancillary data that we have for each of our 4 fields together with the observational depth both in terms of the 1 sigma sky noise as AB magnitude. We characterize the sky noise by placing $\sim500$ apertures, each with a radius of 5 pixels, in sky regions of the image and calculating the RMS of the measured fluxes. 

\subsection{Data Reduction}

The data were obtained from the MAST and reduced by the standard pipeline to obtain rate files. The rate files were further processed using \texttt{Grizli} \citep{brammer2019grizli, brammer2023} to produce combined frames for direct imaging, and GR150R + GR150C dispersed images for each visit, using the \texttt{AstroDrizzle} routine of \texttt{DrizzlePac} \citep{gonzaga2012, hoffmann2021}. 

\subsection{Emission Line Detection}\label{sect:emline_detection}
We develop a source detection algorithm to identify galaxies by their line emission in the dispersed spectral images using the following procedure.
\begin{enumerate}
    \item We create a source detection image by (one dimensional) unsharp masking. We create a smoothed image by running a rolling median filter (width of 30px) along each pixel row in the dispersion direction, and subtracting it from the spectral image. The spectral traces are slightly curved in practice, but this curvature is significantly less than 1 px along the first order, so we do not account for this curvature.  The median filtering efficiently removes continuum light but leaves emission lines as point sources in the subtracted image.
    \item We run \texttt{Source Extractor} \citep{bertin1996} as implemented in the python library \texttt{sep} \citep{barbary2016} on the median filtered image and create a catalog of potential emission lines. We use a \texttt{thresh} of 3 and \texttt{minarea} of 5 which yields a very permissive candidate list.
    \item Next we create a catalog of bright sources in the direct image, again with \texttt{Source Extractor}. From this catalog we calculate the positions of expected zero orders, which we then cross correlate with our emission line catalog (from Step 2). Sources that fall within 5 pixel radius of an expected zero order are excluded from the emission line catalog.
    \item We then visually inspect the direct images, 2D spectra, and extracted 1D spectra of the emission line catalog sources. We classify them as artifacts, low-$z$ emission line sources (usually with multiple emission lines such as [OII], [OIII], or H$\alpha$) or single line emitter candidates. With the very relaxed source extractor settings, we have approximately 800-1000 candidate sources per NIRISS field. Approximately 80-90\% of these are artifacts, commonly from poorly masked hot pixels, or artifacts produced by the median filtering close to the edges of bright spectra.
\end{enumerate}
This procedure can in principle use any of the three NIRISS bands as the detection band, but we only use the F115W since it covers Ly$\alpha$ emission between $z\sim7.5$ and $z\sim9.5$. Future work will extend this search to the longer wavelength bands.

\subsection{Calibration and Extraction}\label{sec: spectrum extraction}

\subsubsection{Wavelength Calibration}
Traditional grism extraction starts from detections in direct imaging and then use this source position to model the source spectrum trace and the positions of surrounding objects to model the potential source contamination. However, we need to search for very faint line-only sources, with varying relative depths in the direct and dispersed images, and in some cases even with missing F115W direct images. Thus, not all sources detected in emission lines will be detected in the direct image. We therefore use spectral images dispersed in both the R and C directions to validate sources and determine spatial location of sources for wavelength calibration. Our procedure is: 
\begin{enumerate}
    \item From the emission line detection location in the initial dispersed image we calculate all possible origin positions in the direct image. This equates to one position per assumed observed wavelength of the line. We refer to this region as the Possible Origin Line (POL).
    \item Given the POL in the direct image we can calculate the possible locations at which the emission line must fall in the second grism orient.
    \item We then perform do source detection along this line in the second grism allowing offsets of maximum of 2.5 pixels on either side of the line to account for possible calibration offsets and shifts because of irregular morphologies. 
    \item Finally, we trace this second line detection back into the direct image and find the point at which the implied POLs intersect.  This gives us the position that matches the line wavelengths and also coordinate we use for the final trace calculations and extractions.
\end{enumerate}

In order to model the trace we use the \texttt{grismconf} formalism as implemented in the \texttt{GRIZLI} code \citep{brammer2019grizli} together with the wavelength calibration files derived  by \citet{pirzkal2023}, who report a wavelength calibration uncertainty of less than half a pixel. 

\subsubsection{Spectral Extraction and Background Subtraction}
We perform our spectral extractions in combined drizzled images produced using the \texttt{GRIZLI} preprocessing steps. The spectral extraction is done using an optimal extraction formalism, meaning that the spectrum is weighted in the cross-dispersion direction by a combination of the PSF size and the size of the source. The source size is measured using a Gaussian fit in the cross dispersion direction at the position of the detected emission line. However we find that all LAE candidates are consistent with being unresolved or marginally resolved, so the extraction weighting is simply the PSF profile, which we approximate as a Gaussian with standard deviation 0.6 pixels.

Due to both contamination by nearby sources and an uneven background in the NIRISS data, local backgrounds can be significant. We therefore do a relatively simple local background subtraction. For each pixel along the predicted trace position we select two 3-pixel regions outside 2 sigma of the source width in the cross dispersion direction. We then fit a straight line to these points to model the background at the trace pixel. This background is then subtracted from the data before the weighted extraction. We find that this simplistic approach in general performs well, but if the contaminating source spectrum contains strong emission lines, it can lead to over-subtraction of the spectrum. We flag/remove such cases during visual inspection.

\subsection{LAE Candidate Validation}\label{sec: validation}

In order to securely identify single line emitters as real sources, we require that the emission line is detected in both grisms at $>2\sigma$ confidence, with  line strengths and widths in both grisms consistent within approximately 50\%. For parallels with deep F115W direct images (Par28 and NGDEEP), we also consider a direct source detection to lend further credence to the source, but for Par50, Par52 we do not have direct F115W imaging. 

Single line emitters in our data most likely have two primary source populations: H$\alpha$ emission from  $z\sim0.7$ and Ly$\alpha$ from  $z \geq 7.5$. Other strong lines like [OII]$_{3727,29}$ and [OIII]$_{5007}$ would also show other strong emission lines in the redder bands.  Bright H$\alpha$-emitters may also be distinguished by other lines such as [SII]$_{6717,31}$, [SIII]$_{9069,9531}$, HeI$_{1.087}$, but detection of these much fainter lines cannot be guaranteed. Differentiating these populations is crucial. The primary differentiation between the populations is the expected lack of any continuum emission blueward of a Ly$\alpha$ line. The faintness of the sources means that continuum is often undetected in the spectra. In order to attempt to lend confidence to our candidates we search for archival data from both HST and ground-based telescopes, primarily Subaru, in filters blueward of F115W. For all the parallels considered here we have collected deep SuprimeCam imaging from Subaru from the Suprime-Cam Legacy Archive \citep{gwyn2020}, and for Par28, we also have deep HST imaging from the Cosmic Evolution Survey \citep[COSMOS][]{scoville2007,Koekemoer2007}. 

To quantify the depth of the ancillary imaging and confirm detections and non-detections of our LAE candidates we need accurate noise level estimates. We obtain this for each ancillary image by measuring the fluxes in 3 pixel apertures placed in empty regions of the image and then calculating the RMS of these measurements (right columns of Table~\ref{tab:observations}). We then measure the flux at the expected source location in a matched aperture and require that flux to be below the 1 $\sigma$ sky level.

\begin{figure*}
    \centering
    \includegraphics[width=\linewidth]{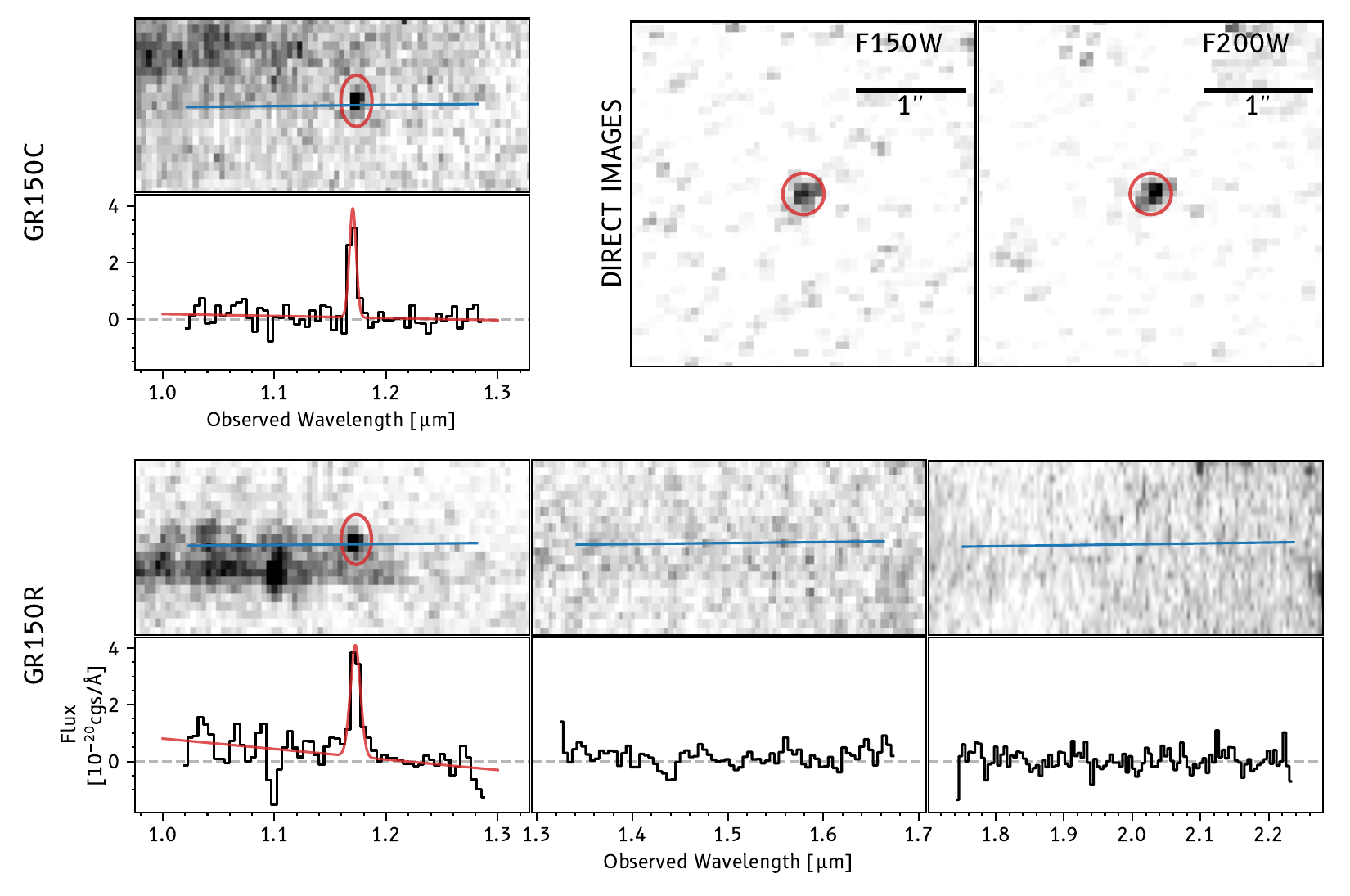}
    \caption{Data for \textbf{P50CLAE1}. The top left panels show the G150C 2D and 1D spectra. The blue line indicates the central position of the trace and the red ellipse highlights the approximate position of the emission line. The bottom rows show the same for the G150R grism. For this grism all three filters are shown in the columns. Extracted 1D spectra of the source are shown in F$_\lambda \times 10^{20}$. The top right panels show the direct image cutouts of the source from the F150W and F200W filters together with a red circle indicating the source coordinate used for extraction.}
    \label{fig:Par50-CLAE1}
\end{figure*}

\begin{figure*}
    \centering
    \includegraphics[ width=\linewidth]{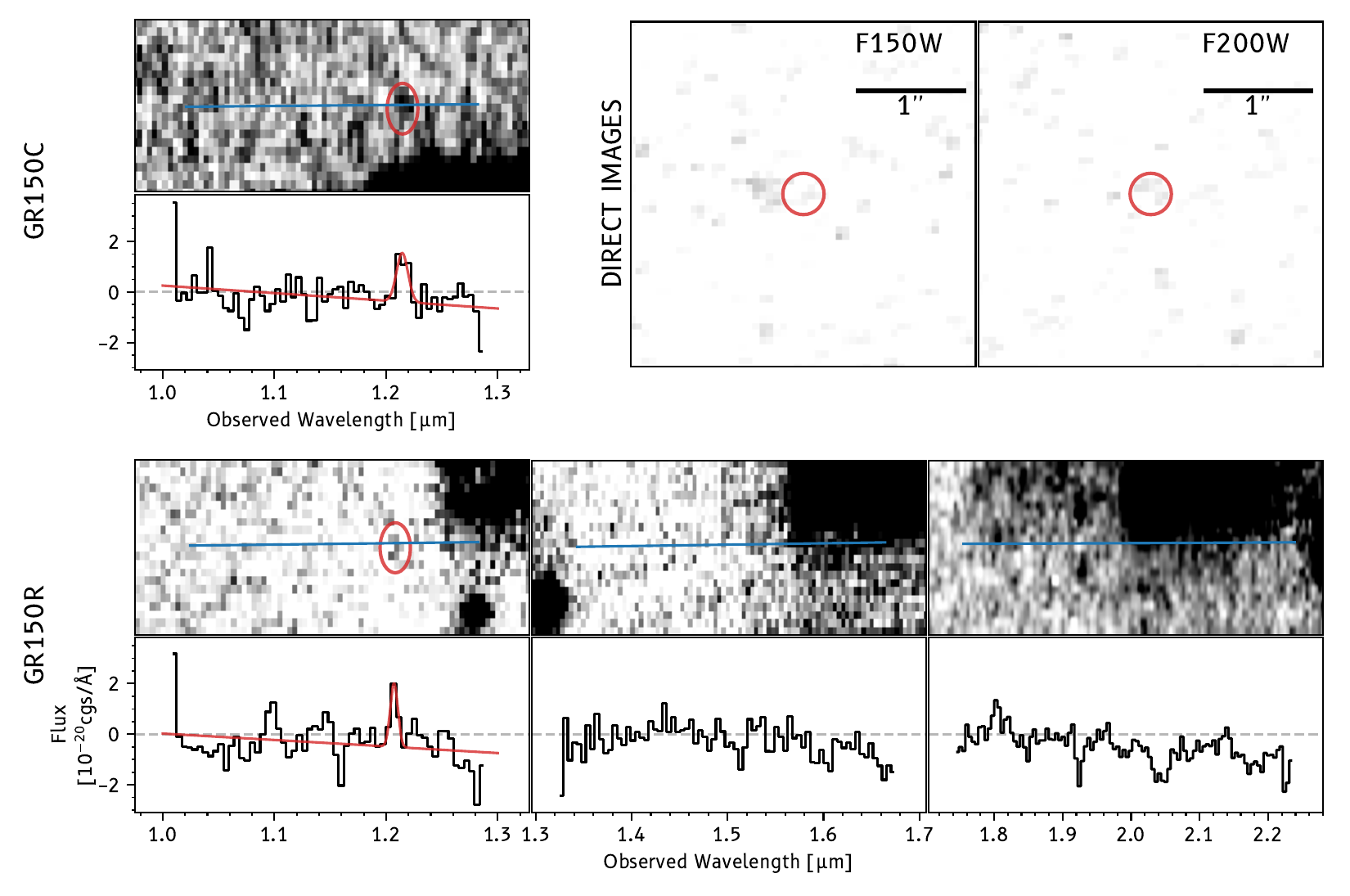}
    \caption{Same as Figure~\ref{fig:Par50-CLAE1} but for P50CLAE2}
    \label{fig:P50CLAE2}
\end{figure*}

\begin{figure*}
    \centering    \includegraphics[width=\linewidth]{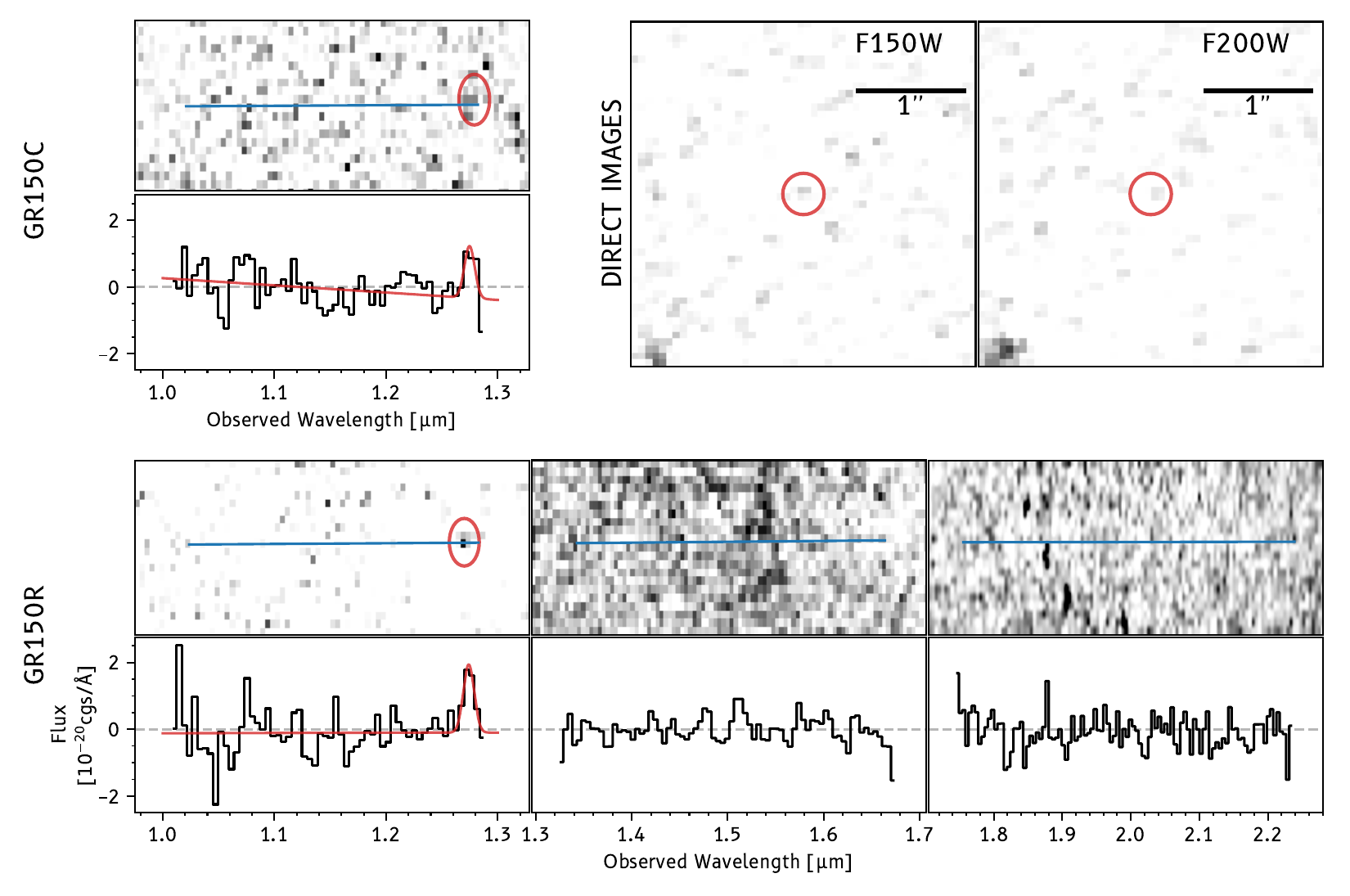}
    \caption{Same as Figure~\ref{fig:Par50-CLAE1} but for P50CLAE3}
    \label{fig:P50CLAE3}
\end{figure*}

\begin{figure*}
    \centering    \includegraphics[width=\linewidth]{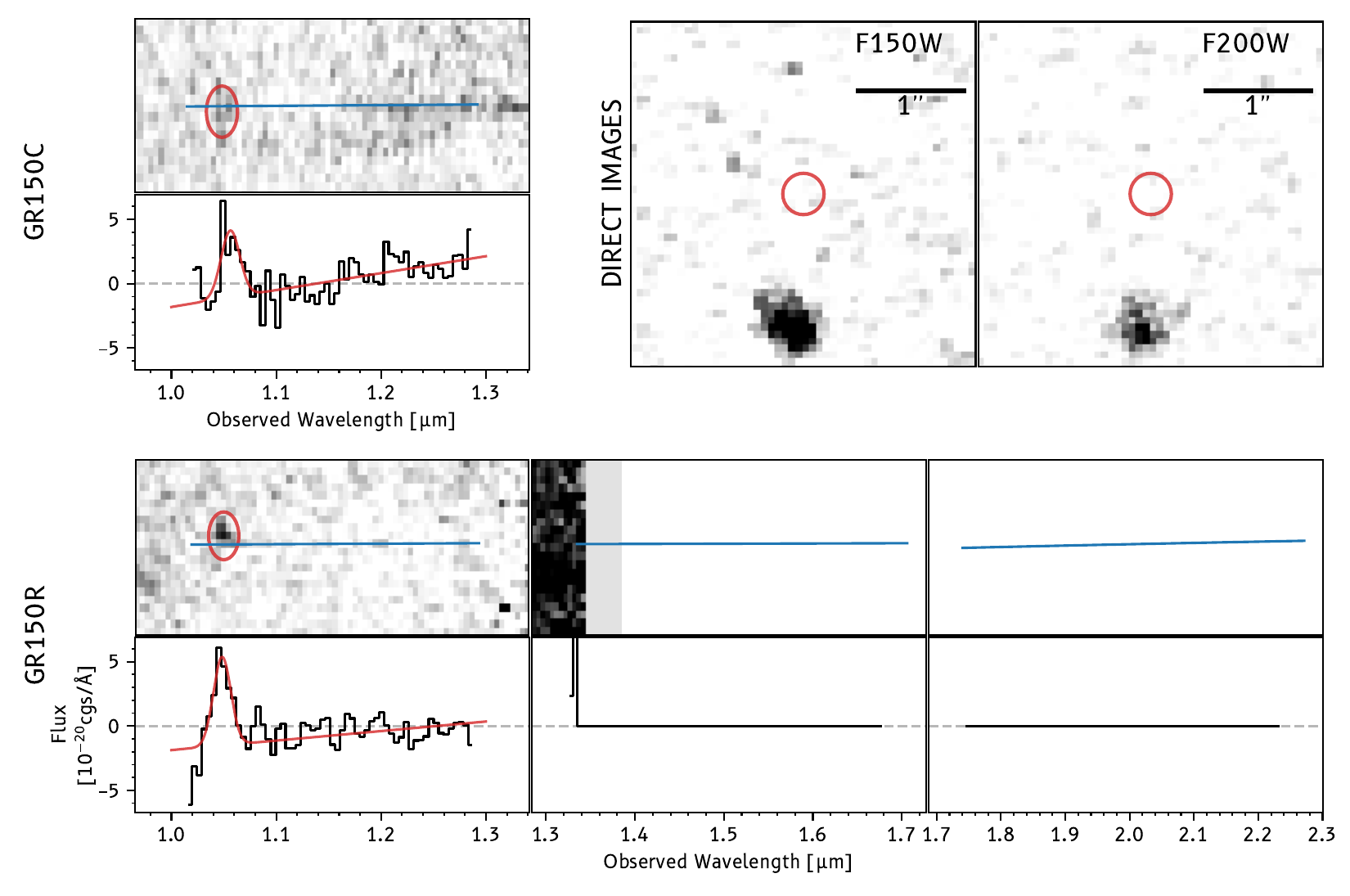}
    \caption{Same as Figure~\ref{fig:Par50-CLAE1} but for P52CLAE1}
    \label{fig:P52CLAE1}
\end{figure*}

\section{LAE candidate sample}\label{sec:sample}
After candidate selection and validation we have reduced a sample of $\sim 4000$ initial selections to a final sample of 4 candidate line emitters with redshifts between 7.6 and  9.5. In this section we present measurements of these emitters as well as the implied populations statistics. 

The direct images, 2D spectra, and extracted 1D spectra are shown in Figures\,\ref{fig:Par50-CLAE1} to \ref{fig:P50CLAE3}. First we present how we characterize the sources together with some discussion of each individual candidate, and then  discuss sample properties, such as completeness and implied luminosity functions.

\subsection{Source Characterization}

\begin{deluxetable*}{lccccccc}
\tabletypesize{\scriptsize}
\tablewidth{0pt} 
\tablecaption{Observed Ly$\alpha$ characteristics  \label{tab:results}}
\tablehead{
\colhead{Source} & \colhead{RA} & \colhead{Dec} & \colhead{Redshift} & \colhead{Luminosity$^1$} & \colhead{Flux$^2$}& \colhead{Width$^{3,4}$}  & \colhead{EW$^{4,5}$}}
\colnumbers
\startdata 
P50CLAE1 & 189.18718 & 62.05699 & 8.63$\pm0.004$ & $3.4\pm0.4$  & 36.9$\pm4.4$ & 30$\pm6$   & 70.1 \\
P50CLAE2 & 189.15394 & 62.06494 & 9.01$\pm0.01$  & $1.5\pm0.3$  & 15.3$\pm3$   & 52$\pm8$   & 36.9 \\ 
P50CLAE3 & 189.16159 & 62.07098 & 9.48$\pm0.02$  & $2.8\pm0.45$ & 24.7$\pm4$   & 47$\pm11$  & 45.1 \\
P52CLAE1 & 150.17060 & 2.03315  & 7.60$\pm0.01$  & $8.4\pm0.8$  & 122.4$\pm12$ & 88.5$\pm3$ & 98.6 \\
\enddata
\tablecomments{$^1$In units of $10^{42}$erg/s. $^2$In units of $10^{-19}$erg/s/cm$^2$. $^3$1$\sigma$ line width. $^4$In units of Å. $^5$Restframe 1$\sigma$ lower limit. }
\end{deluxetable*}

To measure the line flux, we fit the extracted spectra with single Gaussian line profiles for each dispersion direction separately. While a simple background subtraction is done during the extraction, some residual background can remain, which we account for by simultaneously fitting a linear component.  Line measurements are then averaged over the two grism orients to produce final fluxes, central wavelengths, and line widths. 

Comparing the uncertainty frames produced by the pipeline with the observed pixel-to-pixel fluctuations we find that the pipeline uncertainty in general appears to significantly underestimate the true uncertainty.
For this reason we instead estimate the noise level of a given spectrum by measuring the standard deviation of all pixels, excluding a wavelength window around the emission line set to be 2 $\sigma_\mathrm{line}$ wide, after subtracting a linear background to make sure no trends will skew the measurements. We then use a Monte Carlo (MC) procedure to estimate the final flux errors: we randomize the spectra by drawing flux perturbations from a Gaussian with mean zero and standard deviation equal to the fluctuations measured in the original spectrum, and refit the emission line. This is repeated 1000 times and the final errors are calculated as the standard deviation of the resulting distribution of flux measurements.

\begin{figure}
    \centering
    \includegraphics[width=\linewidth]{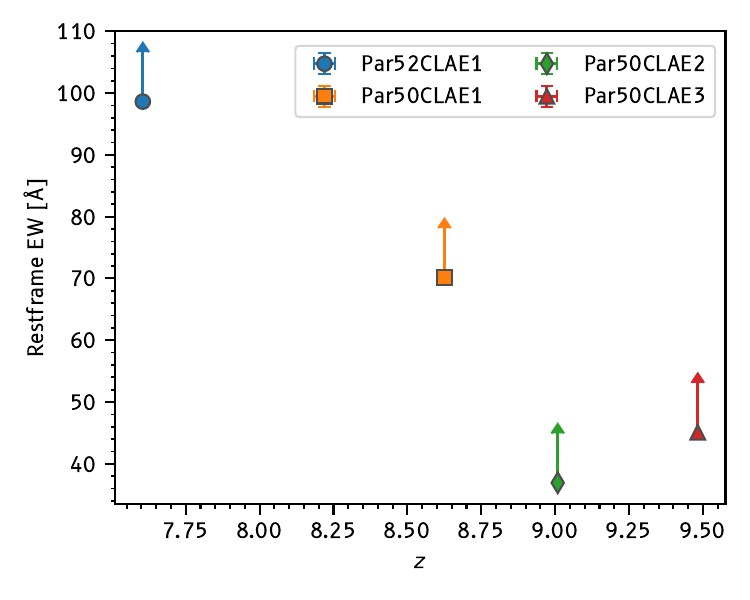}
    \caption{Equivalent width limits for the four LAE candidates as a function of redshift. All points are lower 1$\sigma$ limits since the continuum is undetected.}
    \label{fig:ews}
\end{figure}

We do not detect continuum emission in the spectra of any of our LAEs. Despite this we can still place limits on the EWs. We do this simply by considering a maximum continuum level equal to the local noise measured from the spectrum. The resulting lower limits on the EWs are shown in Figure~\ref{fig:ews}. 
We find rest-frame equivalent widths above $\sim$30 Å for all sources. This is expected from the resolution of NIRISS, are is comparable to the commonplace definitions of Ly$\alpha$ emitters used in the literature \citep[e.g.][]{stark2011, pentericci2011, napolitano2024a}.

\subsubsection{P50CLAE1}
P50CLAE1 is shown in Figure\,\ref{fig:Par50-CLAE1}. This is the clearest candidate of the sample and has an unambiguous line detection at 11710Å, corresponding to a redshift of 8.63. The line is narrow, with a line width of only 30Å (corresponding to $\approx770$km/s, noting that 1 NIRISS pixel corresponds to $\approx1200$km/s at this redshift), and is detected at more than 12$\sigma$ significance despite the noticeable contamination from a nearby source in the GR150R grism image. The source is also visible, although faint,  in both the F150W and F200W images. 

\subsubsection{P50CLAE2}
This source, shown in Figure\,\ref{fig:P50CLAE2} is the faintest of our candidates, has a redshift of 9.01, and a rest-frame EW of 36.9 Å. The line falls close to strong contaminants in both dispersion directions but fortunately remains in a region where it can be detected. The source is undetected in the imaging in F150W and F200W (see upper right images in Figure\,\ref{fig:P50CLAE2}). Such non-detections are to be expected, since the exposure times of the direct images are only one third of the dispersed exposure time, and no strong emission lines are expected in the wavelength regions covered by the F150W and F200W when the Ly$\alpha$ line falls in the F115W filter.  This highlights the importance of searching for line emission in the dispersed images, as well as direct imaging. 

\subsubsection{P50CLAE3}
In Figure\,\ref{fig:P50CLAE3} we show the highest redshift candidate in the sample, which, at $z=9.48$, lies at the very edge of the F115W bandpass. The line is very weak, but nevertheless corresponds to  a restframe EW of $\approx45$Å. We also do not detect the source in the F150W or F200W direct images, for the same reasons as described for P50CLAE2 above. 

\subsubsection{P52CLAE1}
This is the lowest redshift source, shown in Figure\,\ref{fig:P52CLAE1}, and the most ambiguous of our detections. The reason is that the source lies close to the edge of the field which means that the F150W and F200W spectra fall outside the edge of the dispersed image. Additionally the line detected is significantly broader than the other Ly$\alpha$ detections and the source has a cross dispersion size which is larger than PSF by an additional $\sigma=1.4$ pixels. The source remains undetected in F150W and F200W direct imaging.

\subsection{Completeness estimation}
\begin{figure*}
    \centering
    \includegraphics[width=1\linewidth]{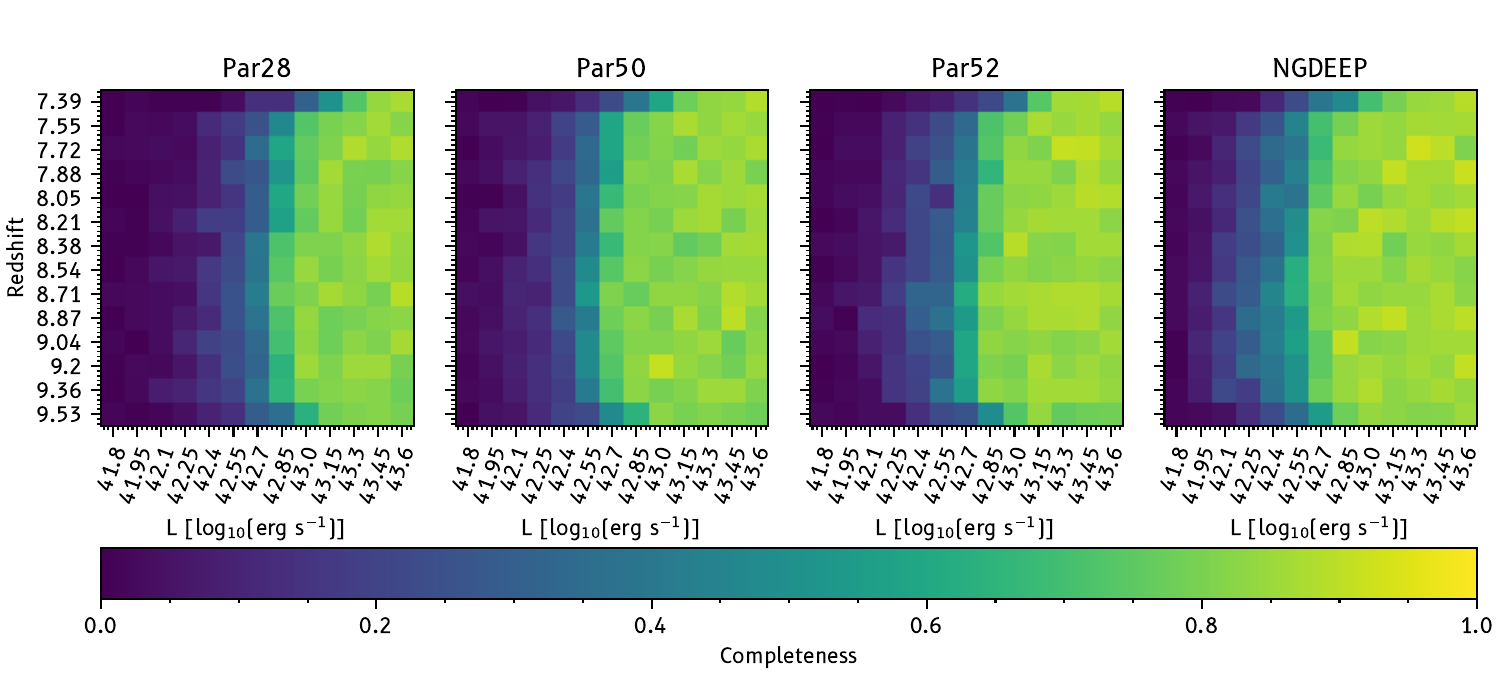}
    \caption{Completeness levels for each field as a function of redshift and luminosity (log$_{10}$(ergs$^{-1}$)). Trends like the decrease of completeness close to the edges of the filter(corresponding to low and high redshifts) can be clearly seen.}
    \label{fig:completeness}
\end{figure*}

In order to determine accurate number densities and  luminosity functions (see Section\,\ref{sec:lfs}) we need a precise understanding of our sample completeness. We estimate the completeness using a source injection and recovery simulation following \citet{herenz2019}. The simulation mimics the source selection procedure as closely as possible but, requiring the simulations to be automated, we therefore cannot include the visual inspection steps of the procedure. 

Since all our candidates all are either unresolved or marginally resolved, we simplify our simulations by limiting them to point-sources. We model the sources as pure emission line sources i.e. a given line flux, an observed wavelength of the line, and zero continuum. The emission line is modeled as a Gaussian line profile of PSF width (FWHM$=1.42$px) in both the dispersion and cross dispersion directions (corresponding to 0.094'' in the spatial direction and 1700\,km\,s$^{-1}$ in the dispersion direction, assuming the average redshift of our sample). From the generated synthetic 1D spectrum we compute the inverse of the optimal extraction weighting along a trace  calculated for a randomly chosen point in the direct frame. This produces a 2D spectrum which we then add to the science frame. When new source locations are chosen we make sure that no sources are placed within 15 pixels of each other in order to avoid artificial crowding.

The new seeded frame is then passed to source detection where we do median filtering and source extraction using the same settings as for the actual source detection. For each source that is recovered in the GR150C dispersion direction we then extract the GR150R spectrum and do line fitting on the known location of the simulated line, as we would on the real detections. The noise in the spectrum is estimated from the spectral regions outside a 2 sigma wide window around the emission line and used as input to a 500 iteration Monte Carlo simulation of the line fitting. This allows us to estimate the error on the line detection. We then consider a source rediscovered in both dispersion directions if the fitted line has a higher signal to noise than $2\sigma$.

We run these simulations for luminosities between $10^{41.8}$ and  $10^{43.5}$ erg\,s$^{-1}$ in steps of $\log(L)=0.15$, and $7.4\leq z\leq9.5$ in steps of 0.16, and for each combination we generate 3 images of 150 sources each. This sampling allows us to construct a grid in L and $z$ which we can sample to get the completeness for a given source. This simulation procedure is done for each field independently. 

We present the results of the completeness simulations in Figure\,\ref{fig:completeness}. We can note a few interesting features.
The first is that the completeness for all fields reaches a maximum around 80-90\%. This is because we have to mask a 200-pixel wide region along the border of the dispersed images, in which the zeroth order images of sources that lie outside the direct image are located.  These mimic emission lines, but we cannot model and exclude them using the direct image (see Section~\ref{sect:emline_detection}). This is not the quite same as standard completeness, however the reduction in effective volume is the same when used in conjunction with the $C^{-}$ and $1/V_\mathrm{max}$ methods (see Section\,\ref{sec:lfs}). The second noteworthy feature is the relative depths of the fields. We note that NGDEEP is the deepest field as expected, but that the depths of Par50, 52 and 28 are comparable and Par28 may indeed be the least complete of the four fields, despite having the longest exposure time of the PASSAGE parallels. This is due to unfortunate bright star contamination of that field. 

\subsection{Luminosity Functions}\label{sec:lfs}
One of the most fundamental properties of a galaxy population is its LF. The LAE LF has been found to be approximately constant up to $z\sim6$ \citep{ouchi2008, herenz2019}, above which a precipitous drop occurs that is often interpreted as the impact of an increasingly neutral IGM \citep{kashikawa2011}. To date, all LAEs above $z=7.3$  have been selected by their continuum emission (e.g. as LBGs) and similar. Here we instead have an, albeit small, sample of line-selected Ly$\alpha$ emitting galaxies. It is therefore interesting to examine the implied LF of these galaxies and how it differs from other selection techniques. 

\begin{figure}
    \centering
    \includegraphics[width=1\linewidth]{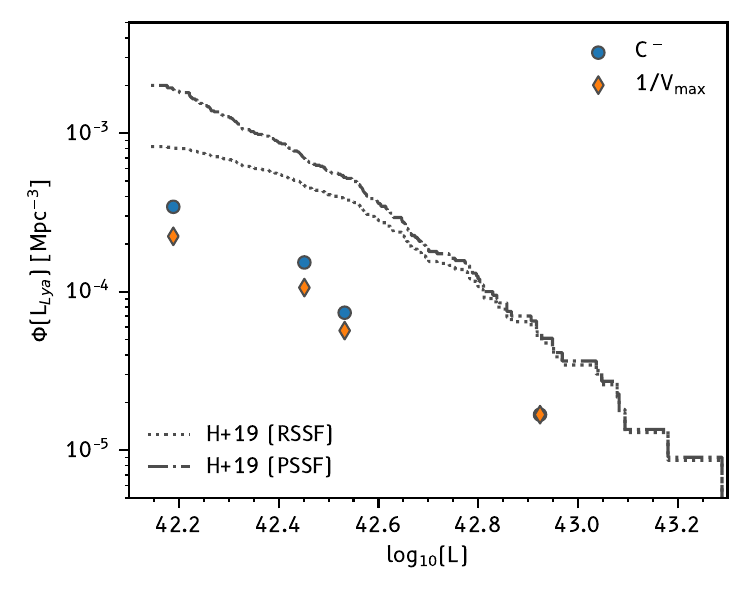}
    \caption{Cumulative Ly$\alpha$ luminosity function derived with the C$^-$ method (blue points) and 1/V$_\mathrm{max}$ (orange points). The black lines (dotted and dot-dashed) are the LFs determined by \citet{herenz2019} (H+19) between redshift $\sim3$ and $\sim6$ with the dot-dashed line using a completeness function that considers LAEs as point sources and the dotted line considering them as spatially extended. }
    \label{fig:lyalf}
\end{figure}

We determine the cumulative luminosity function $\Phi$ of our sample using two different methodologies: the $1/V_\mathrm{max}$ method and the $C^-$ method.  The $1/V_\mathrm{max}$ method is a refinement of a simple number count per total observed volume, and calculates the number density by considering the maximum volume within which a specific source could be detected, given the completeness simulation presented in the last section. $C^-$ on the other hand calculates the cumulative LF using a product of Dirac-$\delta$ functions. We refer the reader to \citet{johnston2011} and \citet{herenz2019} for more detailed discussions of these methods. We present the two resulting LFs in Figure\,\ref{fig:lyalf} together with an LF of $3\lesssim z \lesssim 6.6$ LAEs from \citet{herenz2019}, noting that our observed densities are significantly lower than those at lower $z$. 

The small number of sources prevents us from deriving a full differential luminosity  function, $\phi$, usually parametrized by the Schechter function \citep{Schechter1976}. However, we can examine the relative normalization by assuming the same Schechter shape parameters as redshift 6 and scaling the function to fit our points. We derive $\phi$ in two luminosity bins, with the sample divided below and above $10^{42.5}$\,erg\,s$^{-1}$, resulting in two galaxies in each bin. We then rescale the Schechter fit from \citet{herenz2019} to optimally fit our points by recalculating the normalization $\phi_*$.

\begin{figure}
    \centering
    \includegraphics[width=1\linewidth]{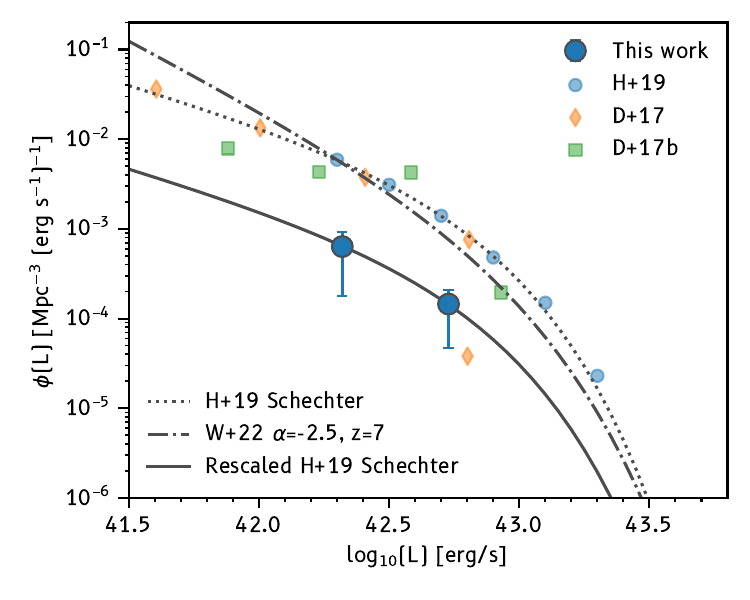}
    \caption{Differential luminosity function. blue points indicate data from this work using the binned $1/V_{max}$ estimator. Dotted line indicates the global LAE LF from \citet{herenz2019} derived for redshift 3 to 6.7 (see Figure\ref{fig:lyalf}), and the solid line is the \citet{herenz2019} LF rescaled to fit our measurements. Dot-dashed line indicates the LF from \citet{wold2022}. Green diamonds and red squares indicate LF measurements from \citet{Drake2017} and \citet{Drake2017b} respectively at $2.91 < z < 6.64$.}
    \label{fig:diff_lf}
\end{figure}

The results are presented in Figure\,\ref{fig:diff_lf} and the rescaled $\mathrm{log}_{10}(\phi_*)$ value is -3.64 as compared to -2.71 from \citet{herenz2019}. We can see that the rescaled $\phi_*$ value lies significantly below the \citet{herenz2019} result by a factor of $\sim$10, indicating a considerable reduction in the volume density of LAEs. This downwards evolution in the LF is expected for a universe that is significantly more neutral at $z=8$ than it is at $z=6$ and only Ly$\alpha$ emission from galaxies residing in large ionized regions will be observable.

\section{Implications for Reionization}\label{sec:reionization}

Here we examine the implications that our observations and LFs have on the the neutral content of the Universe, paying specific attention to \lya\ EWs, transmission through the IGM, and the comparison of empirical and simulated LFs.

    \subsection{Intrinsic Equivalent Widths}
    \begin{figure}
        \centering
        \includegraphics[width=\linewidth]{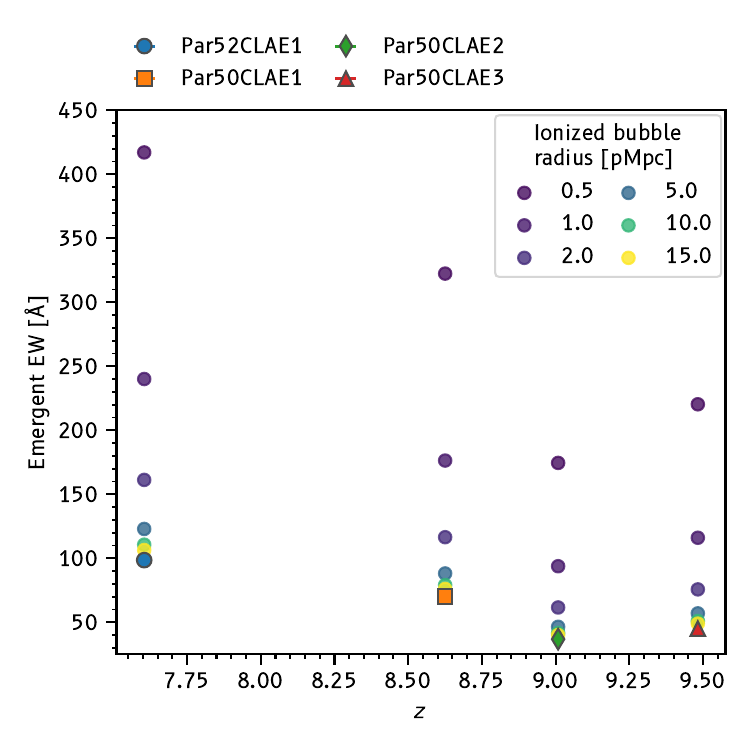}
        \caption{Implied emergent EWs for our sources as a function of the size of the ionized bubble the galaxy resides in.}
        \label{fig:EW extrapolation}
    \end{figure}

    The EWs of LAEs have been used in many studies to place constraints on the reionization process \citep[see e.g.][]{nakane2024}. While we only have lower limits on the EW of our emitters, we can still use these to estimate the limits of the rest-frame EW of the galaxy before IGM absorption. As in \citet{hayesscarlata2023}, we refer to this as the `emergent equivalent width'. In order to calculate the emergent EW of a source, we have to make a few key assumptions. First we assume that the universe is largely neutral at these redshifts, but that the galaxies lie in ionized regions. Second, we assume an intrinsic Ly$\alpha$ spectral profile, for which we use the median stack of COS observations of low-z Ly$\alpha$ emitting galaxies from \citet{hayes2021} since the constituent galaxies in the stack are at sufficiently low redshift that IGM absorption is negligible. This low-$z$ profile is observationally indistinguishable from spectral profiles observed out to at least $z=5.5$.
    Third, we require a  radius of the ionized bubble that the galaxy resides in, which we vary between 0.5 and 15 pMpc. 
    
    In order to calculate the IGM attenuation at the redshifts of our galaxy for a given ionized bubble size we follow the methodology of \citet{witstok2024b}, and make use of the code \texttt{lymana\_absorption}\footnote{\url{https://github.com/joriswitstok/lymana\_absorption}}. We calculate the IGM transmission curve as a function of wavelength which we use to calculate the fraction of transmitted light from our assumed Ly$\alpha$ profile. We then multiply our rest-frame EWs with this scaling factor to get the emergent EW. The results are shown in Figure\,\ref{fig:EW extrapolation}.

    We note that only moderately large ionized bubbles are required for the emergent EWs to be in line with theoretical bounds and observed LAE EWs. Requiring the maximum emergent EW to be less than 250Å would only require bubbles $\gtrsim 2$~pMpc. This is broadly consistent with the radii reported by \citet{hayesscarlata2023} and \citet{umeda2024}.  However, 250Å is not a hard limit but simply a typical value for a zero-age stellar population with a Salpeter IMF, and galaxies at these redshifts likely do not conform to these assumptions. For instance there have been several suggestions that as we approach very low metallicities, the IMF may become top heavy \citep[see e.g.][for some examples]{hayes2024, Cameron2024, nandal2024, hutter2024}. One should also note that this calculation assumes an escape fraction of Ly$\alpha$ of 1. A more realistic $f_{esc}$ assumption would require larger bubble sizes to remain below a given maximum theoretical EW. However the observed emergent rest-frame EWs themselves imply a high $f_{esc}$ in these galaxies \citep[e.g.][]{sobral2019}.
    
    \subsection{Ly$\alpha$ Transmission Factor}
    Following \citet{wold2022} we now try to quantify the evolution of the Ly$\alpha$ IGM transmission using the Ly$\alpha$ transmission factor, which is defined as 
    $$T_{Ly\alpha}= \frac{\rho_{Ly\alpha}^{z2} / \rho_{Ly\alpha}^{z1}}{\rho_{UV}^{z2}/\rho_{UV}^{z1}},$$ where $\rho_{Ly\alpha}^{z}$ is the Ly$\alpha$ luminosity density at a given redshift $z$ and $\rho_{UV}^{z}$ the UV luminosity density. The luminosity densities are calculated as $\int L\phi(L) dL$  using the Schechter function fit of $\phi$, and we consider integration limits between $M_\mathrm{UV}=-22$ and $-18$ mag, and $\log(L/\mathrm{erg\,s}^{-1})=42.4$ and $45$ for the UV LF and the Ly$\alpha$ LFs, respectively.

    We compute the Ly$\alpha$ density at redshift 7.5--9.5 from  the rescaled Schechter function presented in Section\,\ref{sec:lfs}, and the $z\sim6$ Ly$\alpha$ density from the original \citet{herenz2019} fit. For the UV LF we use the redshift 5.9 function from \citet{bouwens2021} as the low redshift anchor point and \citet{bouwens2023} for the high-$z$ measurement. 
    We find that the implied $T_{Ly\alpha}$ is 0.27, i.e. the transmission at redshift 7.5--9.5, dominated by the $z=8.5-9.5$ end, is a quarter of that at redshift 6. This implies a significant evolution in the IGM neutral fraction between these redshifts, but not as strong as would be expected if reionization was late and rapid \citep[e.g.][]{naidu2020}. We note that a reduction in \lya\ transmission of a factor of four is the same as the factor-of-four drop in \lya\ fraction that \citet{tang2024a} estimate between redshift 5.5 and 8 using \lya\ followup of continuum-selected galaxies. 
    \begin{figure*}
        \centering
        \includegraphics[width=\linewidth]{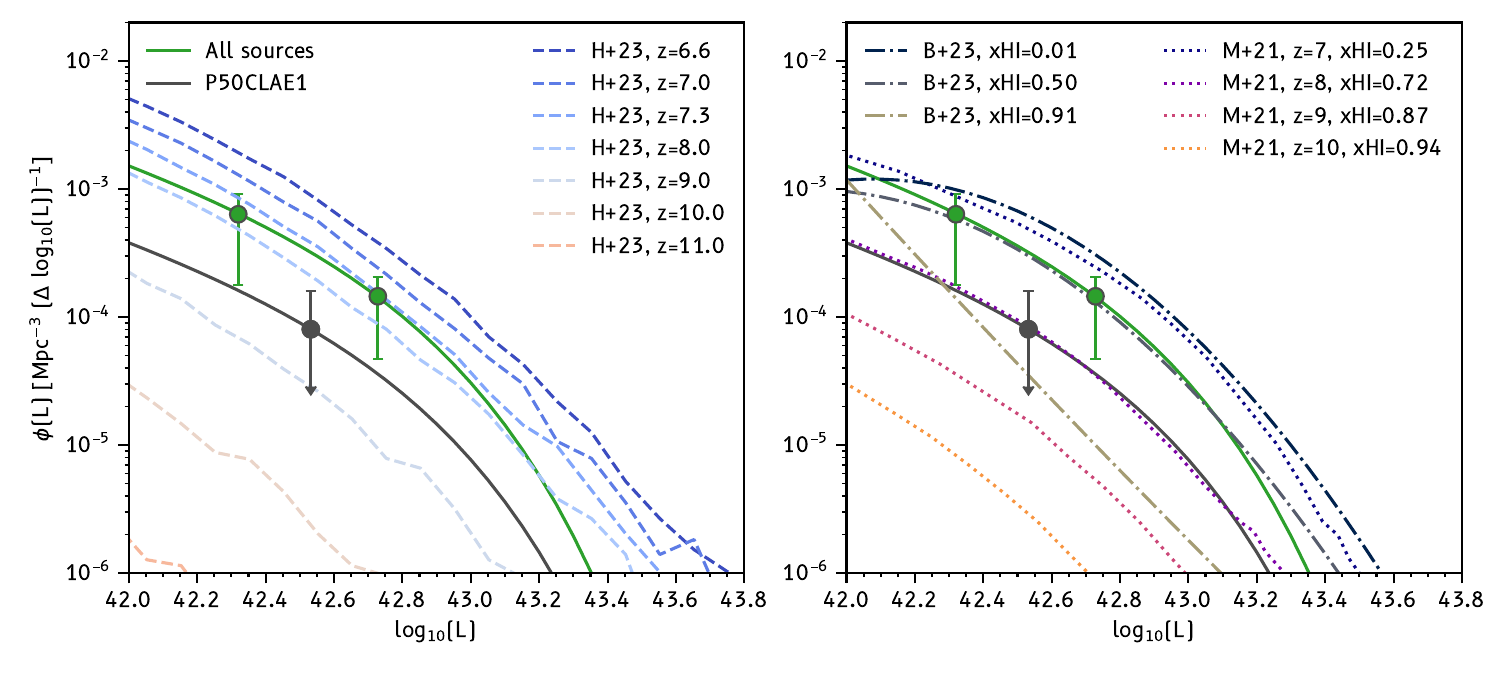}
        \caption{Comparison of the differential LFs derived in this work (green circles), with those from various simulations. 
        Left panel: simulations are taken from the \texttt{ASTRAEUS} simulation of   \citet[][H+23]{hutter2023}, computed for a Salpeter IMF at various redshifts. Right panel: simulations are taken from \citet[][M+21]{morales2021} using dotted lines, and the analytical fit to the LF from \citet[][B+23]{bruton2023} using dot-dashed lines. We also include versions of our LFs where we only consider P50CLAE1, the highest significance source, which is shown as the black circle and solid line.}
        \label{fig:differential_comparison}
    \end{figure*}
    
    One should note two things about this method however: the first is that a transmission factor does not uniquely  correspond to a neutral fraction, since it would depend on the topology of the ionized regions in the IGM. Secondly, this method of calculating the transmission may be oversimplified by assuming a non-evolving EW distribution (although the elapsed cosmic time over this interval is only 350 Myr, and perhaps large changes in the emergent EW distribution should not be expected).

    \subsection{Comparisons to Expected Ly$\alpha$ LFs}\label{sec: comparisons to expectations}
    Another way to use our findings to make inferences on the reionization history is to directly compare them to theoretical predictions. We compare our differential LFs to theoretical predictions in Figure\,\ref{fig:differential_comparison}.

    In the left panel, we contrast the observed LFs with those of \citet{hutter2023, hutter2024}, computed using the \texttt{ASTRAEUS} simulation framework  \citep{hutter2021} which includes both a semi-analytic galaxy formation model and a semi-numeric reionization simulation, assuming a Salpeter IMF. From the figure it is clear that our observed LF would best correspond to the simulated LF at redshift 7.3, and is quite discrepant from the expected level at our mean redshift of 8.6. One should note that there are some potential caveats in the simulations, such as the assumption that the emergent Ly$\alpha$ is a Gaussian profile with no velocity offset, which lower redshift observations show to essentially never be the case \citep{hayes2023}.  \citet{hutter2023} do show that assuming double peaked Ly$\alpha$ makes only marginal difference to the LF. However, there is a possibility that more red peak dominated emergent line profiles, similar to those observed at lower-$z$, would reduce the IGM sensitivity of the Ly$\alpha$ and boost the observed LF. The observed LF would then match simulated LFs at higher redshift, although we unfortunately cannot precisely estimate the redshift of best agreement. 
    
    \citet{hutter2024} also present \texttt{ASTRAEUS} simulations using an evolving IMF, where dense gas produces a top heavy IMF and enhances the number of massive-ionizing stars. Indeed, the IMF is not expected to be constant as we reach early cosmic times, when the ISM of star-forming becomes denser, although the exact nature of the evolution is hard to predict \citep[e.g.][]{Cameron2024}.  A consequence of such an IMF would be that lower mass galaxies produce more ionizing and Ly$\alpha$ photons, resulting in different reionization topology and  Ly$\alpha$ LF \citep{hutter2024}. We also compared our data to the Evolving IMF simulations, and found that our observations then are well matched with the simulated LF at $z\simeq 8$. However, one should note here that the impact of the evolving IMF is relatively small at redshifts below 9, and becomes much more pronounced thereafter, which may also help in explaining the detection of a $z\sim9.5$ source in our sample.
    
    We now turn  to the right panel of Figure\,\ref{fig:differential_comparison} where we plot the LF predictions of \citet{bruton2023} at $z=7$ for three different values of $x_{HI}$, and LFs for $7 \leq z \leq 10$ for the simulated average $x_{HI}$ at those redshifts from \citet{morales2021}. We note that our data points lie in the upper span of the models presented, matching closely the line of \citet{bruton2023} for a neutral fraction of 0.5. Considering the \citet{morales2021} LFs our data lies between the $z=7$ and $z=8$ data points. Again our points fall above the simulated LFs, but are marginally consistent with the $z=8$ expectations at $1\sigma$.

    \subsection{Uncertainties on the Luminosity Function}\label{sect:lf_uncertainties}
    There are three main contributors to the uncertainty on our LF determinations: Poisson/counting statistics in the low-number regime, the unknown interloper/contamination fraction, and cosmic variance (CV). In this subsection we address each of these in turn. 

    We estimate the impact of counting statistics by making some simplifying assumptions: firstly that the luminosity distribution within a LF bin is uniform, secondly that objects do not scatter between bins, and thirdly that the bin widths are static. In order to estimate the counting errors, shown as errorbars in Figure\,\ref{fig:diff_lf} and Figure\,\ref{fig:differential_comparison}, we resample the number of sources in the bin from a Poisson distribution of mean 2.  For each source we then assign a luminosity from a uniform distribution in log($L$) with edges set by the bin edges. These new sources are then used to calculate $\phi$ using the $1/V_\mathrm{max}$ methodology as before. We repeat this procedure 1000 times and show the 16th and 84th percentiles as the asymmetric errorbars in the figures.
    
    Estimating the impact of contamination is non-trivial but to give a conservative measure we omit all sources except for P50CLAE1, which is the highest confidence detection, and recompute the LFs. This assumption is very conservative, but useful for establishing limits of the implied LFs. The result is shown as the black circle and solid line in Figure\,\ref{fig:differential_comparison}. To compute the differential LF we have to make a choice of the bin-width which we set to 0.5. The resulting point is the outcome of one $z=8.6$ galaxy, and now matches with the $z=8$ curve from \citet{morales2021}, significantly reducing the tension with simulations. When comparing to the \texttt{ASTRAEUS} simulations of \citet{hutter2023}, this LF is consistent with expectations between $z\sim8$ and 9. 
    
    CV  describes the scatter in the measurements that is intrinsic to the Universe, encapsulating intrinsic dark matter halo variation in the volume, bias, and the hydrogen-reionization process itself that determines the \lya\ visibility. Pure CV uncertainties can be calculated from the dark matter density field and are largest in small contiguous fields, but both the large $\Delta z$ of our \lya\ survey and the four independent pointings do work to reduce CV.  Using the CV calculator of \citet{trentistiavelli2008}, we calculate an additional CV error budget of only 14\% assuming dark matter halo masses above $10^{10}$\,M$_\odot$.  In order to include the effects of reionization/IGM topology on the LAE counts, we need to turn to a simulation.  We construct coarse light-cones from the simulation of \citet{bruton2023}, using the average reionization history updated for the JWST Cycle 1 LF in \citet[][their Figure~9]{mason2025}.  We then select random locations in the cube, extract regions corresponding to the NIRISS field-of-view and redshift range $z=7.5-9.5$, and count the number of LBGs with $M_\mathrm{UV}<-19$ and LAEs with $L_\mathrm{Ly\alpha}>1.5\times 10^{42}$\,erg\,s$^{-1}$ (the least luminous object in our sample and an exact $5\sigma$ detection; Table~\ref{tab:results}).  We repeat this for four pointings to mimic our survey, and execute a 1000-realization Monte Carlo simulation to predict the number of LBGs and LAEs. At the 16th and 84th percentiles, we expect to observe 4--10 LBGs and 0--3 LAEs.  With 4 LAEs in our survey, we slightly exceed the expected values, as would also be predicted from our LFs.  We develop these arguments, and compare our result with the \texttt{ASTRAEUS} simulations in Sections~\ref{sec: implications of observations} and \ref{sect:improving}.

\section{Discussion}\label{sec:discussion}

\subsection{Implications of Recent \lya\ Observations}\label{sec: implications of observations}
Pre-JWST results suggested that finding significant numbers of Ly$\alpha$ emitting galaxies at redshifts beyond 7 was going to be very challenging with only a few detections reported \citep{schenker2012, schenker2014, hoag2019, tilvi2020, jung2022}. Large ground-based narrowband surveys, such as UltraVISTA, targeted $z\sim8$ LAEs but found none, due to the difficulty of reaching the required flux limits against the bright atmospheric background in the NIR: the limiting flux of UltraVISTA was about $1.5\times10^{-17}$\,erg\,s$^{-1}$\,cm$^{-2}$, while  our $z>8$ detections have an average flux that is $\simeq 6$ times fainter ($2.4\times10^{-18}$\,erg\,s$^{-1}$\,cm$^{-2}$). However, the situation has drastically changed after the launch of JWST and detections now reach deep into the EoR. Detecting sources like GNz11 \citep{bunker2023}, and now a LAE at $z=13$ \citep{witstok2024b}, was completely unexpected. \citet{tang2024a} used observations of the EGS, GOODS-N, GOODS-S and Abell2744, covering 41 NIRSpec pointings (corresponding to total area coverage $\gtrsim 440$ arcmin$^2$), to detect 33 galaxies with Ly$\alpha$ emission above redshift 6.5, as well as confirm a $7\leq z\leq8$ galaxy overdensity in the EGS field. In this work we detect three LAEs with a similar $\Delta z$ to galaxies in the EGS overdensity, within a single NIRISS pointing of only $\sim4.8$ arcmin$^2$. 

The question that arises is whether the unexpected prevalence of Ly$\alpha$ at high redshift in fact has implications for models of the EoR. In a reionization scenario that is mainly powered by a few bright galaxies (as suggested by for instance \citet{naidu2022}, sometimes referred to as oligarchal reionization) we would expect to detect Ly$\alpha$ emitters in a relatively clustered manner, as indeed we may be doing now, but we would also expect such sightlines to be relatively rare and that the LAEs detected should be dominated by bright sources. Given the total area currently surveyed, it appears LAEs are more common than expected, and the LAEs we detect here are also unusually bright. On the other hand, \citet{qin2024} suggest that a more distributed reionization process, where fainter galaxies play a more important role, will boost the likelihood of detecting fainter LAEs. This scenario is more likely to be in line with our findings. 

\citet{hutter2023} show that the Ly$\alpha$ emitters in the \texttt{ASTRAEUS} simulation are dominated by more massive galaxies residing in larger ionized regions. We therefore conduct a test to see what the likelihood is of observing a conjunction of LAEs like the one we report here in a volume corresponding to a single NIRISS pointing. To do this we analyze the population of LAEs in a $z=9$ snapshot of the simulation: we select  LAEs from the Salpeter IMF simulation with luminosities above $1.5\times10^{42}$ erg/s---approximately the lowest luminosity in our $8.5\leq z\leq 9.5$ sample. Then we bin the simulation volume in 120$^3$ cells to obtain LAE number counts in each cell. For each cell we then count the numbers of LAEs within a volume corresponding to the V$_{\mathrm{max}}$ volume estimate for Par50 centered on that cell.

We find that only $\approx$0.21\% of such volumes have $\geq 3$ sources in them. This suggests that the reionization history and topology produced in the Salpeter IMF version of \texttt{ASTRAEUS} is an unlikely representation of the $z=8.6$ universe, when confronted with our NIRISS observations (under the assumption that all our sources are real). To investigate this further, we conduct the same experiment for the simulations with an evolving IMF. The resulting likelihood of $\approx0.74\%$ three times higher than for the Salpeter IMF, but nevertheless remains unlikely. 

To draw fully quantitative conclusions regarding the implications of our observations for EoR morphology would require detailed modeling that is beyond the scope of this work, as well as followup of the sources. Nevertheless, these results, especially when taken in conjunction with other LAE observations, suggest that the picture of LAEs being very rare deep into the EoR may be incorrect. This may imply a distributed reionization topologies and a boost in the observability of more common,  fainter, LAEs.

\subsection{Field-to-field Variance}
Early simulations by \citet{jensen2013} assessed the possibility of using the field-to-field variation in \lya\ emitting galaxies as a discriminator between reionization.  
Indeed it has been found that at least one of the JWST deep fields, EGS, does cover a high redshift overdensity \citep{tilvi2020, jung2022, tang2024a}. Another indication of significant field-to-field variance comes from \citet{napolitano2024a} who measured the Ly$\alpha$ fraction in the CEERS and JADES surveys using MSA spectroscopic data. Their results indicate large differences (up to a factor of 3) between the fields, and depending on whether or not known overdensities are included. The results also appear to cast some doubt over the previously established increase of $X_{Ly\alpha}$ up to redshift 6 and subsequent drop, instead showing a large scatter of values across all redshifts. One should note, however, that the small slitlets of the NIRSpec MSA may lead to large slit losses of Ly$\alpha$, which is typically more extended than the UV emission \citep{hayes2013, wisotzki2016, leclercq2020, runnholm2023}. 

The principal way to overcome and characterize cosmic variance is to observe many spatially uncorrelated fields \citep[e.g.][for \lya\ stuies]{bruton2023}, which indeed was one of the primary motivations for the PASSAGE project. The number counts of LAEs in our fields is quite uneven, with zero LAEs recovered in the deepest NGDEEP data, but three found in one of the shallower parallel fields. The sample dispersion ($\equiv\frac{\mbox{Sample Mean}}{\mbox{Sample Variance}}$) is relatively large (1.5) compared to the expectation (1) for a Poisson process. We therefore test whether this is consistent with being simply a random draw from a uniform Poisson distribution or whether additional cosmic variance is required to explain the data. 

To do this we draw samples of four random variates from a Poisson distribution with the same mean as our observed sample (=1) and calculate the dispersion. We repeat this 10000 times and then compare our observed dispersion to the distribution. We find that the probability of finding a dispersion $\geq 1.5$ is 0.13 i.e. unlikely but not impossible. From this we conclude that while some impact of cosmic variance is probable, we do not require it to explain our observations. Characterizing cosmic variance and the deviations from simple Poissonian statistics in more detail requires more independent field observations. The Cycle 3 program \#3383 (PI: Glazebrook) obtained pure parallel NIRISS data similar to PASSAGE, however, only 2 fields in this program have the required depth and F115W grism coverage to be usable for this purpose.

\subsection{Improving $x_{HI}$ Precision with JWST/NIRISS}\label{sect:improving}

    Obviously more data is required to make more precise estimates of the neutral fraction from Ly$\alpha$ LFs, and here we address the question of how many pointings would be needed. We noted in Section~\ref{sec: comparisons to expectations} that there are three sources of variation on the number counts: cosmic variance (including the overlayed state of the IGM/H{\sc ii} bubble size), interlopers/contamination, and Poisson uncertainty.  Here we assess the Poisson errors directly in our data (e.g. assuming the LF values are accurate but their errorbars are larger than desired), and the impact of CV+reionization on \lya\ visibility using updates to the simulations of \citet[][]{bruton2023}.
    
    The right panel of Figure~\ref{fig:differential_comparison} shows that the high-luminosity green data-point (log($L_\mathrm{Ly\alpha}$/erg\,s$^{-1})=42.72$) provides rather strong constraints on the $x_{HI}$, because the simulated LFs of \citet{bruton2023} diverge at high $L$ (see also \citealt{hu2019} and \citealt{wold2022}). Taking Bruton et al's LFs as truth and assuming purely Poissonian errors, we estimate that reducing the error on $x_{HI}$ to $\Delta x_\mathrm{HI} = 0.1$ would require only 8 fields, or approximately doubling the survey compared to that presented here.  However, the same simulations can also be used to address the CV error budget as in Section~\ref{sect:lf_uncertainties}. With only 8 fields the inference would remain dominated by CV, with estimated uncertainty of $\simeq 70$\% on the LF datapoint, due mainly to the reionization topology.

    However, the high-$L$ datapoint is strongly dependent on the luminosity of a single galaxy, P52CLAE1.  This galaxy is $\simeq 3$ times more luminous than the other three sources and has by far the highest EW (see Table~\ref{tab:results}), making it the least representative object in our sample. The recovery of that galaxy may have been an improbable occurrence (as suggested by the LF alone and field-to-field variations at high luminosity; \citealt{wold2022}), so we repeat the same calculation for the more common low $L_\mathrm{Ly\alpha}$ galaxies in the bin near log($L_\mathrm{Ly\alpha}$/erg\,s$^{-1})=42.3$ (lower green datapoint). In this case the uncertainty of $\pm0.1$ on  $x_{HI}$ can be reached with $\simeq 40$ fields.  With 40 random fields, the CV+reionization budget would add just 0.11~dex to the $\Delta\phi$ uncertainty in the LF, and would become negligible (again following \citealt{bruton2023}). Such a survey would be a 10-fold increase in size compared to this study.
    
    Finally, we address the issue of possible contamination in the same way, using the LF data point derived under the assumption that only one source is real (black point in the right panel of Figure~\ref{fig:differential_comparison}). As an upper limit, only the higher error bound is finite, so we simply estimate the number of required sources (of constant luminosity) required to bring this this upper bound down to the level of $\Delta x_\mathrm{HI}=0.1$.  To reach this level of precision a similar NIRISS campaign would require $\simeq 70$ random, uncorrelated pointings. The CV contribution to the uncertainty on $\phi$ would again be negligible (0.08~dex). 

\section{Summary \& Conclusions}\label{sec:conclusions}

We have conducted what is (to our knowledge) the first blind, unbiased survey for \lya\ emitting galaxies at $z>7.5$ with JWST. 
We have conducted a wide-field slitless spectroscopic survey using the NIRISS instrument, obtained under the Parallel Application of Slitless Spectroscopy for Galaxy Evolution program (PASSAGE).  We focus upon three fields for which spectroscopic data have been obtained for at least two hours in both the row and column dispersion directions using the F115W filter,  spectra in the F150W and F200W filters have also been obtained, and for which deep imaging exists at wavelengths bluewards of the WFSS data.  
We further included data from the NGDEEP program in the Hubble Ultradeep Field, which has obtained similar NIRISS data.  Here we summarize the results presented in this work, and the main conclusions we draw. 
\begin{itemize}
    \item We detect Ly$\alpha$ line emission from 4 LAEs in the redshift range $7.5 \leq z \leq 9.5$ across 4 surveyed NIRISS pointings.  These are the highest redshift \lya-selected galaxies known to date.  Their luminosities fall in the range $1.5$ to $8.4\times 10^{42}$\,erg\,s$^{-1}$.
    \item We determine the lower limit of the restframe \lya\ EWs to be 36 to 99\AA, and investigate what these EWs imply for the emergent EW  given a range of ionized bubble sizes. We conclude that relatively small bubbles of radius $\simeq2$ pMpc would yield physically probable emergent EWs. 
    \item We construct the cumulative and differential \lya\ luminosity functions and contrast them with the low redshift results of \citet{herenz2019}. We find that the LF undergoes significant evolution between redshift 6 and our mean redshift of 8.6. However, this amounts to a factor of $\lesssim 10$, and contrasting this observed LF with theoretical expectations from \citet{hutter2023}, \citet{morales2021} and \citet{bruton2023} shows  the evolution to be significantly less than expected. 
    \item Contrasting the evolution of the Ly$\alpha$ LF to the UV LF from \citet{bouwens2023} indicates that the Ly$\alpha$ LF undergoes significantly more change than the UV LF---demonstrating that it is unlikely that galaxy evolution is the main cause of Ly$\alpha$ LF change and that it is rather due to an increasingly neutral IGM. 
    \item One potential risk in making inferences on the average evolution of a galaxy population based on a small number of fields is the impact of cosmic variance on the results. We find large variations between our observed fields that may indicate that cosmic variance plays a role, however, only 4 fields is not sufficient for us to conclude that the distribution is non-Poissonian. 
    \item We also estimate the increase in the number of observed JWST/NIRISS fields required to reach a precision of $\Delta x_\mathrm{HI}=0.1$. We find that this number ranges from $\sim8$ to $\sim70$ fields depending on whether we can reach the more constraining higher luminosity sources, and the fraction of contamination in our current sample. This would be possible to achieve with future PASSAGE-like pure parallel JWST/NIRISS surveys.
\end{itemize}

\begin{acknowledgements}
This work is based on observations made with the NASA/ESA/CSA James Webb Space Telescope. The data were obtained from the Mikulski Archive for Space Telescopes at the Space Telescope Science Institute, which is operated by the Association of Universities for Research in Astronomy, Inc., under NASA contract NAS 5-03127 for JWST. These observations are associated with programs \#1571 and \#2079.

MJH is supported by the Swedish Research Council (Vetenskapsr\aa{}det) and is Fellow of the Knut \& Alice Wallenberg Foundation. 
MB acknowledges support from the ERC Grant FIRSTLIGHT and Slovenian national research agency ARIS through grants N1-0238 and P1-0188. 
AA is supported by European Union--NextGenerationEU RFF M4C2 1.1 PRIN 2022 project 2022ZSL4BL INSIGHT.
ASL acknowledges support from Knut and Alice Wallenberg Foundation.
HA acknowledge support from CNES, focused on the JWST mission, and the Programme National Cosmology and Galaxies (PNCG) of CNRS/INSU with INP and IN2P3, co-funded by CEA and CNES.
AH acknowledges support by the VILLUM FONDEN under grant 37459. The Cosmic Dawn Center (DAWN) is funded by the Danish National Research Foundation under grant DNRF140. 
We thank the anonymous referee, whose thoughtful comments have enhanced the quality of the manuscript. 
\end{acknowledgements}

\bibliographystyle{aasjournal}

\begin{thebibliography}{}
\expandafter\ifx\csname natexlab\endcsname\relax\def\natexlab#1{#1}\fi
\providecommand{\url}[1]{\href{#1}{#1}}
\providecommand{\dodoi}[1]{doi:~\href{http://doi.org/#1}{\nolinkurl{#1}}}
\providecommand{\doeprint}[1]{\href{http://ascl.net/#1}{\nolinkurl{http://ascl.net/#1}}}
\providecommand{\doarXiv}[1]{\href{https://arxiv.org/abs/#1}{\nolinkurl{https://arxiv.org/abs/#1}}}

\bibitem[{{Atek} {et~al.}(2010){Atek}, {Malkan}, {McCarthy}, {Teplitz}, {Scarlata}, {Siana}, {Henry}, {Colbert}, {Ross}, {Bridge}, {Bunker}, {Dressler}, {Fosbury}, {Martin}, \& {Shim}}]{atek2010}
{Atek}, H., {Malkan}, M., {McCarthy}, P., {et~al.} 2010, \apj, 723, 104, \dodoi{10.1088/0004-637X/723/1/104}

\bibitem[{{Atek} {et~al.}(2024){Atek}, {Labb{\'e}}, {Furtak}, {Chemerynska}, {Fujimoto}, {Setton}, {Miller}, {Oesch}, {Bezanson}, {Price}, {Dayal}, {Zitrin}, {Kokorev}, {Weaver}, {Brammer}, {Dokkum}, {Williams}, {Cutler}, {Feldmann}, {Fudamoto}, {Greene}, {Leja}, {Maseda}, {Muzzin}, {Pan}, {Papovich}, {Nelson}, {Nanayakkara}, {Stark}, {Stefanon}, {Suess}, {Wang}, \& {Whitaker}}]{atek24}
{Atek}, H., {Labb{\'e}}, I., {Furtak}, L.~J., {et~al.} 2024, \nat, 626, 975, \dodoi{10.1038/s41586-024-07043-6}

\bibitem[{Bagley {et~al.}(2017)Bagley, Scarlata, Henry, Rafelski, Malkan, Teplitz, Dai, Baronchelli, Colbert, Rutkowski, Mehta, Dressler, McCarthy, Bunker, Atek, Garel, Martin, Hathi, \& Siana}]{bagley2017}
Bagley, M.~B., Scarlata, C., Henry, A., {et~al.} 2017, The Astrophysical Journal, 837, 11, \dodoi{10.3847/1538-4357/837/1/11}

\bibitem[{Bagley {et~al.}(2024)Bagley, Pirzkal, Finkelstein, Papovich, Berg, Lotz, Leung, Ferguson, Koekemoer, Dickinson, Kartaltepe, Kocevski, Somerville, Yung, Backhaus, Casey, Castellano, Ch{\'a}vez~Ortiz, Chworowsky, Cox, Dav{\'e}, Davis, {Estrada-Carpenter}, Fontana, Fujimoto, Gardner, Giavalisco, Grazian, Grogin, Hathi, Hutchison, Jaskot, Jung, Kewley, Kirkpatrick, Larson, Matharu, Natarajan, Pentericci, {P{\'e}rez-Gonz{\'a}lez}, Ravindranath, Rothberg, Ryan, Shen, Simons, Snyder, Trump, \& Wilkins}]{bagley2024}
Bagley, M.~B., Pirzkal, N., Finkelstein, S.~L., {et~al.} 2024, The Astrophysical Journal Letters, 965, L6, \dodoi{10.3847/2041-8213/ad2f31}

\bibitem[{Barbary(2016)}]{barbary2016}
Barbary, K. 2016, The Journal of Open Source Software, 1, 58, \dodoi{10.21105/joss.00058}

\bibitem[{{Battisti} {et~al.}(2024){Battisti}, {Bagley}, {Rafelski}, {Baronchelli}, {Dai}, {Henry}, {Atek}, {Colbert}, {Malkan}, {McCarthy}, {Scarlata}, {Siana}, {Teplitz}, {Alavi}, {Boyett}, {Bunker}, {Gardner}, {Hathi}, {Masters}, {Mehta}, {Rutkowski}, {Shahinyan}, {Sunnquist}, \& {Wang}}]{battisti2024}
{Battisti}, A.~J., {Bagley}, M.~B., {Rafelski}, M., {et~al.} 2024, \mnras, 530, 894, \dodoi{10.1093/mnras/stae911}

\bibitem[{{Becker} {et~al.}(2024){Becker}, {Bolton}, {Zhu}, \& {Hashemi}}]{becker2024}
{Becker}, G.~D., {Bolton}, J.~S., {Zhu}, Y., \& {Hashemi}, S. 2024, \mnras, 533, 1525, \dodoi{10.1093/mnras/stae1918}

\bibitem[{Bertin \& Arnouts(1996)}]{bertin1996}
Bertin, E., \& Arnouts, S. 1996, Astronomy and Astrophysics Supplement Series, 117, 393, \dodoi{10.1051/aas:1996164}

\bibitem[{{Bolan} {et~al.}(2022){Bolan}, {Lemaux}, {Mason}, {Brada{\v{c}}}, {Treu}, {Strait}, {Pelliccia}, {Pentericci}, \& {Malkan}}]{bolan2022}
{Bolan}, P., {Lemaux}, B.~C., {Mason}, C., {et~al.} 2022, \mnras, 517, 3263, \dodoi{10.1093/mnras/stac1963}

\bibitem[{Bosman {et~al.}(2022)Bosman, Davies, Becker, Keating, Davies, Zhu, Eilers, D'Odorico, Bian, Bischetti, Cristiani, Fan, Farina, Haehnelt, Hennawi, Kulkarni, Mesinger, Meyer, Onoue, Pallottini, Qin, {Ryan-Weber}, Schindler, Walter, Wang, \& Yang}]{bosman2022}
Bosman, S. E.~I., Davies, F.~B., Becker, G.~D., {et~al.} 2022, Monthly Notices of the Royal Astronomical Society, 514, 55, \dodoi{10.1093/mnras/stac1046}

\bibitem[{Bouwens {et~al.}(2023)Bouwens, Illingworth, Oesch, Stefanon, Naidu, {van Leeuwen}, \& Magee}]{bouwens2023}
Bouwens, R., Illingworth, G., Oesch, P., {et~al.} 2023, Monthly Notices of the Royal Astronomical Society, 523, 1009, \dodoi{10.1093/mnras/stad1014}

\bibitem[{Bouwens {et~al.}(2021)Bouwens, Oesch, Stefanon, Illingworth, Labb{\'e}, Reddy, Atek, Montes, Naidu, Nanayakkara, Nelson, \& Wilkins}]{bouwens2021}
Bouwens, R.~J., Oesch, P.~A., Stefanon, M., {et~al.} 2021, The Astronomical Journal, 162, 47, \dodoi{10.3847/1538-3881/abf83e}

\bibitem[{{Brammer}(2019)}]{brammer2019grizli}
{Brammer}, G. 2019, {Grizli: Grism redshift and line analysis software}, Astrophysics Source Code Library, record ascl:1905.001

\bibitem[{Brammer(2023)}]{brammer2023}
Brammer, G. 2023, Grizli, Zenodo, \dodoi{10.5281/ZENODO.8370018}

\bibitem[{Bruton {et~al.}(2023{\natexlab{a}})Bruton, Lin, Scarlata, \& Hayes}]{bruton2023}
Bruton, S., Lin, Y.-H., Scarlata, C., \& Hayes, M.~J. 2023{\natexlab{a}}, The Astrophysical Journal Letters, 949, L40, \dodoi{10.3847/2041-8213/acd5d0}

\bibitem[{Bruton {et~al.}(2023{\natexlab{b}})Bruton, Scarlata, Haardt, Hayes, Mason, Morales, \& Mesinger}]{bruton2023a}
Bruton, S., Scarlata, C., Haardt, F., {et~al.} 2023{\natexlab{b}}, The Astrophysical Journal, 953, 29, \dodoi{10.3847/1538-4357/acd179}

\bibitem[{Bunker {et~al.}(2023)Bunker, Saxena, Cameron, Willott, {Curtis-Lake}, Jakobsen, Carniani, Smit, Maiolino, Witstok, Curti, D'Eugenio, Jones, Ferruit, Arribas, Charlot, Chevallard, Giardino, {de Graaff}, Looser, L{\"u}tzgendorf, Maseda, Rawle, Rix, Del~Pino, Alberts, Egami, Eisenstein, Endsley, Hainline, Hausen, Johnson, Rieke, Rieke, Robertson, Shivaei, Stark, Sun, Tacchella, Tang, Williams, Willmer, Baker, Baum, Bhatawdekar, Bowler, Boyett, Chen, Circosta, Helton, Ji, Kumari, Lyu, Nelson, Parlanti, Perna, Sandles, Scholtz, Suess, Topping, {\"U}bler, Wallace, \& Whitler}]{bunker2023}
Bunker, A.~J., Saxena, A., Cameron, A.~J., {et~al.} 2023, Astronomy and Astrophysics, 677, A88, \dodoi{10.1051/0004-6361/202346159}

\bibitem[{{Bunker} {et~al.}(2024){Bunker}, {Cameron}, {Curtis-Lake}, {Jakobsen}, {Carniani}, {Curti}, {Witstok}, {Maiolino}, {D'Eugenio}, {Looser}, {Willott}, {Bonaventura}, {Hainline}, {{\"U}bler}, {Willmer}, {Saxena}, {Smit}, {Alberts}, {Arribas}, {Baker}, {Baum}, {Bhatawdekar}, {Bowler}, {Boyett}, {Charlot}, {Chen}, {Chevallard}, {Circosta}, {DeCoursey}, {de Graaff}, {Egami}, {Eisenstein}, {Endsley}, {Ferruit}, {Giardino}, {Hausen}, {Helton}, {Hviding}, {Ji}, {Johnson}, {Jones}, {Kumari}, {Laseter}, {L{\"u}tzgendorf}, {Maseda}, {Nelson}, {Parlanti}, {Perna}, {Rauscher}, {Rawle}, {Rix}, {Rieke}, {Robertson}, {Rodr{\'\i}guez Del Pino}, {Sandles}, {Scholtz}, {Sharpe}, {Skarbinski}, {Stark}, {Sun}, {Tacchella}, {Topping}, {Villanueva}, {Wallace}, {Williams}, \& {Woodrum}}]{Bunker2024A&A}
{Bunker}, A.~J., {Cameron}, A.~J., {Curtis-Lake}, E., {et~al.} 2024, \aap, 690, A288, \dodoi{10.1051/0004-6361/202347094}

\bibitem[{{Cameron} {et~al.}(2024){Cameron}, {Katz}, {Witten}, {Saxena}, {Laporte}, \& {Bunker}}]{Cameron2024}
{Cameron}, A.~J., {Katz}, H., {Witten}, C., {et~al.} 2024, \mnras, 534, 523, \dodoi{10.1093/mnras/stae1547}

\bibitem[{Cuby {et~al.}(2007)Cuby, Hibon, Lidman, Le~F{\`e}vre, Gilmozzi, Moorwood, \& {van der Werf}}]{cuby2007}
Cuby, J.~G., Hibon, P., Lidman, C., {et~al.} 2007, Astronomy and Astrophysics, 461, 911, \dodoi{10.1051/0004-6361:20066349}

\bibitem[{{Dewdney} {et~al.}(2009){Dewdney}, {Hall}, {Schilizzi}, \& {Lazio}}]{Dewdney09}
{Dewdney}, P.~E., {Hall}, P.~J., {Schilizzi}, R.~T., \& {Lazio}, T.~J.~L.~W. 2009, IEEE Proceedings, 97, 1482, \dodoi{10.1109/JPROC.2009.2021005}

\bibitem[{Dijkstra(2014)}]{dijkstra2014}
Dijkstra, M. 2014, Publications of the Astronomical Society of Australia, 31, e040, \dodoi{10.1017/pasa.2014.33}

\bibitem[{Dijkstra \& Wyithe(2012)}]{dijkstra2012}
Dijkstra, M., \& Wyithe, J. S.~B. 2012, Monthly Notices of the Royal Astronomical Society, 419, 3181, \dodoi{10.1111/j.1365-2966.2011.19958.x}

\bibitem[{{Drake} {et~al.}(2017{\natexlab{a}}){Drake}, {Guiderdoni}, {Blaizot}, {Wisotzki}, {Herenz}, {Garel}, {Richard}, {Bacon}, {Bina}, {Cantalupo}, {Contini}, {den Brok}, {Hashimoto}, {Marino}, {Pell{\'o}}, {Schaye}, \& {Schmidt}}]{Drake2017}
{Drake}, A.~B., {Guiderdoni}, B., {Blaizot}, J., {et~al.} 2017{\natexlab{a}}, \mnras, 471, 267, \dodoi{10.1093/mnras/stx1515}

\bibitem[{{Drake} {et~al.}(2017{\natexlab{b}}){Drake}, {Garel}, {Wisotzki}, {Leclercq}, {Hashimoto}, {Richard}, {Bacon}, {Blaizot}, {Caruana}, {Conseil}, {Contini}, {Guiderdoni}, {Herenz}, {Inami}, {Lewis}, {Mahler}, {Marino}, {Pello}, {Schaye}, {Verhamme}, {Ventou}, \& {Weilbacher}}]{Drake2017b}
{Drake}, A.~B., {Garel}, T., {Wisotzki}, L., {et~al.} 2017{\natexlab{b}}, \aap, 608, A6, \dodoi{10.1051/0004-6361/201731431}

\bibitem[{{Eisenstein} {et~al.}(2023){Eisenstein}, {Willott}, {Alberts}, {Arribas}, {Bonaventura}, {Bunker}, {Cameron}, {Carniani}, {Charlot}, {Curtis-Lake}, {D'Eugenio}, {Endsley}, {Ferruit}, {Giardino}, {Hainline}, {Hausen}, {Jakobsen}, {Johnson}, {Maiolino}, {Rieke}, {Rieke}, {Rix}, {Robertson}, {Stark}, {Tacchella}, {Williams}, {Willmer}, {Baker}, {Baum}, {Bhatawdekar}, {Boyett}, {Chen}, {Chevallard}, {Circosta}, {Curti}, {Danhaive}, {DeCoursey}, {de Graaff}, {Dressler}, {Egami}, {Helton}, {Hviding}, {Ji}, {Jones}, {Kumari}, {L{\"u}tzgendorf}, {Laseter}, {Looser}, {Lyu}, {Maseda}, {Nelson}, {Parlanti}, {Perna}, {Pusk{\'a}s}, {Rawle}, {Rodr{\'\i}guez Del Pino}, {Sandles}, {Saxena}, {Scholtz}, {Sharpe}, {Shivaei}, {Silcock}, {Simmonds}, {Skarbinski}, {Smit}, {Stone}, {Suess}, {Sun}, {Tang}, {Topping}, {{\"U}bler}, {Villanueva}, {Wallace}, {Whitler}, {Witstok}, \& {Woodrum}}]{Eisenstein2023arXiv}
{Eisenstein}, D.~J., {Willott}, C., {Alberts}, S., {et~al.} 2023, arXiv e-prints, arXiv:2306.02465, \dodoi{10.48550/arXiv.2306.02465}

\bibitem[{Finkelstein {et~al.}(2019)Finkelstein, D'Aloisio, Paardekooper, {Russell Ryan Jr.}, Behroozi, Finlator, Livermore, Sanderbeck, Vecchia, \& Khochfar}]{finkelstein2019}
Finkelstein, S.~L., D'Aloisio, A., Paardekooper, J.-P., {et~al.} 2019, The Astrophysical Journal, 879, 36, \dodoi{10.3847/1538-4357/ab1ea8}

\bibitem[{Finkelstein {et~al.}(2023)Finkelstein, Bagley, Ferguson, Wilkins, Kartaltepe, Papovich, Yung, Arrabal~Haro, Behroozi, Dickinson, Kocevski, Koekemoer, Larson, Le~Bail, Morales, {P{\'e}rez-Gonz{\'a}lez}, Burgarella, Dav{\'e}, Hirschmann, Somerville, Wuyts, Bromm, Casey, Fontana, Fujimoto, Gardner, Giavalisco, Grazian, Grogin, Hathi, Hutchison, Jha, Jogee, Kewley, Kirkpatrick, Long, Lotz, Pentericci, Pierel, Pirzkal, Ravindranath, Ryan, Trump, Yang, Bhatawdekar, Bisigello, Buat, Calabr{\`o}, Castellano, Cleri, Cooper, Croton, Daddi, Dekel, Elbaz, Franco, Gawiser, Holwerda, {Huertas-Company}, Jaskot, Leung, Lucas, Mobasher, Pandya, Tacchella, Weiner, \& Zavala}]{finkelstein2023a}
Finkelstein, S.~L., Bagley, M.~B., Ferguson, H.~C., {et~al.} 2023, The Astrophysical Journal, 946, L13, \dodoi{10.3847/2041-8213/acade4}

\bibitem[{{Gonzaga} {et~al.}(2012){Gonzaga}, {Hack}, {Fruchter}, \& {Mack}}]{gonzaga2012}
{Gonzaga}, S., {Hack}, W., {Fruchter}, A., \& {Mack}, J. 2012, {The DrizzlePac Handbook}

\bibitem[{{Greig} {et~al.}(2024){Greig}, {Mesinger}, {Ba{\~n}ados}, {Becker}, {Bosman}, {Chen}, {Davies}, {D'Odorico}, {Eilers}, {Gallerani}, {Haehnelt}, {Keating}, {Lai}, {Qin}, {Ryan-Weber}, {Satyavolu}, {Wang}, {Yang}, \& {Zhu}}]{grieg2024}
{Greig}, B., {Mesinger}, A., {Ba{\~n}ados}, E., {et~al.} 2024, \mnras, 530, 3208, \dodoi{10.1093/mnras/stae1080}

\bibitem[{Gronke {et~al.}(2015)Gronke, Dijkstra, Trenti, \& Wyithe}]{gronke2015}
Gronke, M., Dijkstra, M., Trenti, M., \& Wyithe, S. 2015, Monthly Notices of the Royal Astronomical Society, 449, 1284, \dodoi{10.1093/mnras/stv329}

\bibitem[{Gwyn(2020)}]{gwyn2020}
Gwyn, S. 2020, 527, 575

\bibitem[{{Hayes} {et~al.}(2011){Hayes}, {Schaerer}, {{\"O}stlin}, {Mas-Hesse}, {Atek}, \& {Kunth}}]{hayes2011}
{Hayes}, M., {Schaerer}, D., {{\"O}stlin}, G., {et~al.} 2011, \apj, 730, 8, \dodoi{10.1088/0004-637X/730/1/8}

\bibitem[{{Hayes} {et~al.}(2013){Hayes}, {{\"O}stlin}, {Schaerer}, {Verhamme}, {Mas-Hesse}, {Adamo}, {Atek}, {Cannon}, {Duval}, {Guaita}, {Herenz}, {Kunth}, {Laursen}, {Melinder}, {Orlitov{\'a}}, {Ot{\'\i}-Floranes}, \& {Sandberg}}]{hayes2013}
{Hayes}, M., {{\"O}stlin}, G., {Schaerer}, D., {et~al.} 2013, \apjl, 765, L27, \dodoi{10.1088/2041-8205/765/2/L27}

\bibitem[{{Hayes} {et~al.}(2021){Hayes}, {Runnholm}, {Gronke}, \& {Scarlata}}]{hayes2021}
{Hayes}, M.~J., {Runnholm}, A., {Gronke}, M., \& {Scarlata}, C. 2021, \apj, 908, 36, \dodoi{10.3847/1538-4357/abd246}

\bibitem[{{Hayes} {et~al.}(2023){Hayes}, {Runnholm}, {Scarlata}, {Gronke}, \& {Rivera-Thorsen}}]{hayes2023}
{Hayes}, M.~J., {Runnholm}, A., {Scarlata}, C., {Gronke}, M., \& {Rivera-Thorsen}, T.~E. 2023, \mnras, 520, 5903, \dodoi{10.1093/mnras/stad477}

\bibitem[{{Hayes} {et~al.}(2024){Hayes}, {Saldana-Lopez}, {Citro}, {James}, {Mingozzi}, {Scarlata}, {Martinez}, \& {Berg}}]{hayes2024}
{Hayes}, M.~J., {Saldana-Lopez}, A., {Citro}, A., {et~al.} 2024, arXiv e-prints, arXiv:2411.09262, \dodoi{10.48550/arXiv.2411.09262}

\bibitem[{{Hayes} \& {Scarlata}(2023)}]{hayesscarlata2023}
{Hayes}, M.~J., \& {Scarlata}, C. 2023, \apjl, 954, L14, \dodoi{10.3847/2041-8213/acee6a}

\bibitem[{Herenz {et~al.}(2019)Herenz, Wisotzki, Saust, Kerutt, Urrutia, Diener, Schmidt, Marino, {de la Vieuville}, Boogaard, Schaye, Guiderdoni, Richard, \& Bacon}]{herenz2019}
Herenz, E.~C., Wisotzki, L., Saust, R., {et~al.} 2019, Astronomy and Astrophysics, 621, A107, \dodoi{10.1051/0004-6361/201834164}

\bibitem[{{Hoag} {et~al.}(2019){Hoag}, {Brada{\v{c}}}, {Huang}, {Mason}, {Treu}, {Schmidt}, {Trenti}, {Strait}, {Lemaux}, {Finney}, \& {Paddock}}]{hoag2019}
{Hoag}, A., {Brada{\v{c}}}, M., {Huang}, K., {et~al.} 2019, \apj, 878, 12, \dodoi{10.3847/1538-4357/ab1de7}

\bibitem[{{Hoffmann} {et~al.}(2021){Hoffmann}, {Mack}, {Avila}, {Martlin}, {Cohen}, \& {Bajaj}}]{hoffmann2021}
{Hoffmann}, S.~L., {Mack}, J., {Avila}, R., {et~al.} 2021, in American Astronomical Society Meeting Abstracts, Vol. 238, American Astronomical Society Meeting Abstracts, 216.02

\bibitem[{{Hu} {et~al.}(2019){Hu}, {Wang}, {Zheng}, {Malhotra}, {Rhoads}, {Infante}, {Barrientos}, {Yang}, {Jiang}, {Kang}, {Perez}, {Wold}, {Hibon}, {Jiang}, {Khostovan}, {Valdes}, {Walker}, {Galaz}, {Coughlin}, {Harish}, {Kong}, {Pharo}, \& {Zheng}}]{hu2019}
{Hu}, W., {Wang}, J., {Zheng}, Z.-Y., {et~al.} 2019, \apj, 886, 90, \dodoi{10.3847/1538-4357/ab4cf4}

\bibitem[{Hutter {et~al.}(2024)Hutter, Cueto, Dayal, Gottl{\"o}ber, Trebitsch, \& Yepes}]{hutter2024}
Hutter, A., Cueto, E.~R., Dayal, P., {et~al.} 2024, Astraeus {{X}}: {{Indications}} of a Top-Heavy Initial Mass Function in Highly Star-Forming Galaxies from {{JWST}} Observations at Z{$>$}10, \dodoi{10.48550/arXiv.2410.00730}

\bibitem[{Hutter {et~al.}(2021)Hutter, Dayal, Yepes, Gottl{\"o}ber, Legrand, \& Ucci}]{hutter2021}
Hutter, A., Dayal, P., Yepes, G., {et~al.} 2021, Monthly Notices of the Royal Astronomical Society, 503, 3698, \dodoi{10.1093/mnras/stab602}

\bibitem[{Hutter {et~al.}(2023)Hutter, Trebitsch, Dayal, Gottl{\"o}ber, Yepes, \& Legrand}]{hutter2023}
Hutter, A., Trebitsch, M., Dayal, P., {et~al.} 2023, Monthly Notices of the Royal Astronomical Society, 524, 6124, \dodoi{10.1093/mnras/stad2230}

\bibitem[{Itoh {et~al.}(2018)Itoh, Ouchi, Zhang, Inoue, Mawatari, Shibuya, Harikane, Ono, Kusakabe, Shimasaku, Fujimoto, Iwata, Kajisawa, Kashikawa, Kawanomoto, Komiyama, Lee, Nagao, \& Taniguchi}]{itoh2018}
Itoh, R., Ouchi, M., Zhang, H., {et~al.} 2018, The Astrophysical Journal, 867, 46, \dodoi{10.3847/1538-4357/aadfe4}

\bibitem[{Jensen {et~al.}(2013)Jensen, Laursen, Mellema, Iliev, {Sommer-Larsen}, \& Shapiro}]{jensen2013}
Jensen, H., Laursen, P., Mellema, G., {et~al.} 2013, Monthly Notices of the Royal Astronomical Society, 428, 1366, \dodoi{10.1093/mnras/sts116}

\bibitem[{{Johnston}(2011)}]{johnston2011}
{Johnston}, R. 2011, \aapr, 19, 41, \dodoi{10.1007/s00159-011-0041-9}

\bibitem[{Jung {et~al.}(2022)Jung, Finkelstein, Larson, Hutchison, Straughn, Bagley, Castellano, Cleri, Cooper, Dickinson, Ferguson, Holwerda, Kartaltepe, Kim, Koekemoer, Papovich, Park, Pentericci, {Perez-Gonzalez}, Song, Tacchella, Weiner, Willmer, \& Zavala}]{jung2022}
Jung, I., Finkelstein, S.~L., Larson, R.~L., {et~al.} 2022, New \$z {$>$} 7\$ {{Lyman-alpha Emitters}} in {{EGS}}: {{Evidence}} of an {{Extended Ionized Structure}} at \$z {\textbackslash}sim 7.7\$, \dodoi{10.48550/arXiv.2212.09850}

\bibitem[{Kashikawa {et~al.}(2011)Kashikawa, Shimasaku, Matsuda, Egami, Jiang, Nagao, Ouchi, Malkan, Hattori, Ota, Taniguchi, Okamura, Ly, Iye, Furusawa, Shioya, Shibuya, Ishizaki, \& Toshikawa}]{kashikawa2011}
Kashikawa, N., Shimasaku, K., Matsuda, Y., {et~al.} 2011, The Astrophysical Journal, 734, 119, \dodoi{10.1088/0004-637X/734/2/119}

\bibitem[{Keating {et~al.}(2020)Keating, Weinberger, Kulkarni, Haehnelt, Chardin, \& Aubert}]{keating2020}
Keating, L.~C., Weinberger, L.~H., Kulkarni, G., {et~al.} 2020, Monthly Notices of the Royal Astronomical Society, 491, 1736, \dodoi{10.1093/mnras/stz3083}

\bibitem[{{Kikuta} {et~al.}(2023){Kikuta}, {Ouchi}, {Shibuya}, {Liang}, {Umeda}, {Matsumoto}, {Shimasaku}, {Harikane}, {Ono}, {Inoue}, {Yamanaka}, {Kusakabe}, {Momose}, {Kashikawa}, {Matsuda}, \& {Lee}}]{kikuta2023}
{Kikuta}, S., {Ouchi}, M., {Shibuya}, T., {et~al.} 2023, \apjs, 268, 24, \dodoi{10.3847/1538-4365/ace4cb}

\bibitem[{{Koekemoer} {et~al.}(2007){Koekemoer}, {Aussel}, {Calzetti}, {Capak}, {Giavalisco}, {Kneib}, {Leauthaud}, {Le F{\`e}vre}, {McCracken}, {Massey}, {Mobasher}, {Rhodes}, {Scoville}, \& {Shopbell}}]{Koekemoer2007}
{Koekemoer}, A.~M., {Aussel}, H., {Calzetti}, D., {et~al.} 2007, \apjs, 172, 196, \dodoi{10.1086/520086}

\bibitem[{{Koopmans} {et~al.}(2015){Koopmans}, {Pritchard}, {Mellema}, {Aguirre}, {Ahn}, {Barkana}, {van Bemmel}, {Bernardi}, {Bonaldi}, {Briggs}, {de Bruyn}, {Chang}, {Chapman}, {Chen}, {Ciardi}, {Dayal}, {Ferrara}, {Fialkov}, {Fiore}, {Ichiki}, {Illiev}, {Inoue}, {Jelic}, {Jones}, {Lazio}, {Maio}, {Majumdar}, {Mack}, {Mesinger}, {Morales}, {Parsons}, {Pen}, {Santos}, {Schneider}, {Semelin}, {de Souza}, {Subrahmanyan}, {Takeuchi}, {Vedantham}, {Wagg}, {Webster}, {Wyithe}, {Datta}, \& {Trott}}]{koopmans2015}
{Koopmans}, L., {Pritchard}, J., {Mellema}, G., {et~al.} 2015, in Advancing Astrophysics with the Square Kilometre Array (AASKA14), 1, \dodoi{10.22323/1.215.0001}

\bibitem[{Kulkarni {et~al.}(2019)Kulkarni, Keating, Haehnelt, Bosman, Puchwein, Chardin, \& Aubert}]{kulkarni2019}
Kulkarni, G., Keating, L.~C., Haehnelt, M.~G., {et~al.} 2019, Monthly Notices of the Royal Astronomical Society, 485, L24, \dodoi{10.1093/mnrasl/slz025}

\bibitem[{{Kusakabe} {et~al.}(2022){Kusakabe}, {Verhamme}, {Blaizot}, {Garel}, {Wisotzki}, {Leclercq}, {Bacon}, {Schaye}, {Gallego}, {Kerutt}, {Matthee}, {Maseda}, {Nanayakkara}, {Pell{\'o}}, {Richard}, {Tresse}, {Urrutia}, \& {Vitte}}]{kusakabe2022}
{Kusakabe}, H., {Verhamme}, A., {Blaizot}, J., {et~al.} 2022, \aap, 660, A44, \dodoi{10.1051/0004-6361/202142302}

\bibitem[{Laursen {et~al.}(2019)Laursen, {Sommer-Larsen}, {Milvang-Jensen}, Fynbo, \& Razoumov}]{laursen2019}
Laursen, P., {Sommer-Larsen}, J., {Milvang-Jensen}, B., Fynbo, J. P.~U., \& Razoumov, A.~O. 2019, Astronomy and Astrophysics, 627, A84, \dodoi{10.1051/0004-6361/201833645}

\bibitem[{Leclercq {et~al.}(2020)Leclercq, Bacon, Verhamme, Garel, Blaizot, Brinchmann, Cantalupo, Claeyssens, Conseil, Contini, Hashimoto, Herenz, Kusakabe, Marino, Maseda, Matthee, Mitchell, Pezzulli, Richard, Schmidt, \& Wisotzki}]{leclercq2020}
Leclercq, F., Bacon, R., Verhamme, A., {et~al.} 2020, Astronomy and Astrophysics, 635, A82, \dodoi{10.1051/0004-6361/201937339}

\bibitem[{{Mason} {et~al.}(2025){Mason}, {Chen}, {Stark}, {Lu}, {Topping}, \& {Tang}}]{mason2025}
{Mason}, C.~A., {Chen}, Z., {Stark}, D.~P., {et~al.} 2025, arXiv e-prints, arXiv:2501.11702, \dodoi{10.48550/arXiv.2501.11702}

\bibitem[{Mason {et~al.}(2018)Mason, Treu, Dijkstra, Mesinger, Trenti, Pentericci, {de Barros}, \& Vanzella}]{mason2018}
Mason, C.~A., Treu, T., Dijkstra, M., {et~al.} 2018, The Astrophysical Journal, 856, 2, \dodoi{10.3847/1538-4357/aab0a7}

\bibitem[{Mason {et~al.}(2019)Mason, Fontana, Treu, Schmidt, Hoag, Abramson, Amorin, Brada{\v c}, Guaita, Jones, Henry, Malkan, Pentericci, Trenti, \& Vanzella}]{mason2019}
Mason, C.~A., Fontana, A., Treu, T., {et~al.} 2019, Monthly Notices of the Royal Astronomical Society, 485, 3947, \dodoi{10.1093/mnras/stz632}

\bibitem[{McCracken {et~al.}(2012)McCracken, {Milvang-Jensen}, Dunlop, Franx, Fynbo, Le~F{\`e}vre, Holt, Caputi, Goranova, Buitrago, Emerson, Freudling, Hudelot, {L{\'o}pez-Sanjuan}, Magnard, Mellier, M{\o}ller, Nilsson, Sutherland, Tasca, \& Zabl}]{mccracken2012}
McCracken, H.~J., {Milvang-Jensen}, B., Dunlop, J., {et~al.} 2012, Astronomy and Astrophysics, 544, A156, \dodoi{10.1051/0004-6361/201219507}

\bibitem[{Morales {et~al.}(2021)Morales, Mason, Bruton, Gronke, Haardt, \& Scarlata}]{morales2021}
Morales, A.~M., Mason, C.~A., Bruton, S., {et~al.} 2021, The Astrophysical Journal, 919, 120, \dodoi{10.3847/1538-4357/ac1104}

\bibitem[{{Naidu} {et~al.}(2020){Naidu}, {Tacchella}, {Mason}, {Bose}, {Oesch}, \& {Conroy}}]{naidu2020}
{Naidu}, R.~P., {Tacchella}, S., {Mason}, C.~A., {et~al.} 2020, \apj, 892, 109, \dodoi{10.3847/1538-4357/ab7cc9}

\bibitem[{Naidu {et~al.}(2022)Naidu, Matthee, Oesch, Conroy, Sobral, Pezzulli, Hayes, Erb, Amor{\'i}n, Gronke, Schaerer, Tacchella, Kerutt, {Paulino-Afonso}, Calhau, Llerena, \& R{\"o}ttgering}]{naidu2022}
Naidu, R.~P., Matthee, J., Oesch, P.~A., {et~al.} 2022, Monthly Notices of the Royal Astronomical Society, 510, 4582, \dodoi{10.1093/mnras/stab3601}

\bibitem[{Nakane {et~al.}(2024)Nakane, Ouchi, Nakajima, Harikane, Ono, Umeda, Isobe, Zhang, \& Xu}]{nakane2024}
Nakane, M., Ouchi, M., Nakajima, K., {et~al.} 2024, The Astrophysical Journal, 967, 28, \dodoi{10.3847/1538-4357/ad38c2}

\bibitem[{{Nandal} {et~al.}(2024){Nandal}, {Sibony}, \& {Tsiatsiou}}]{nandal2024}
{Nandal}, D., {Sibony}, Y., \& {Tsiatsiou}, S. 2024, \aap, 688, A142, \dodoi{10.1051/0004-6361/202348866}

\bibitem[{Napolitano {et~al.}(2024)Napolitano, Pentericci, Santini, Calabr{\`o}, Mascia, Llerena, Castellano, Dickinson, Finkelstein, Amorin, Haro, Bagley, Bhatawdekar, Cleri, Davis, Gardner, Gawiser, Giavalisco, Hathi, Hu, Jung, Kartaltepe, Koekemoer, Merlin, Mobasher, Papovich, Park, Pirzkal, Trump, Wilkins, \& Yung}]{napolitano2024a}
Napolitano, L., Pentericci, L., Santini, P., {et~al.} 2024, Astronomy \& Astrophysics, 688, A106, \dodoi{10.1051/0004-6361/202449644}

\bibitem[{{Oke} \& {Gunn}(1983)}]{oke1983}
{Oke}, J.~B., \& {Gunn}, J.~E. 1983, \apj, 266, 713, \dodoi{10.1086/160817}

\bibitem[{Ono {et~al.}(2012)Ono, Ouchi, Mobasher, Dickinson, Penner, Shimasaku, Weiner, Kartaltepe, Nakajima, Nayyeri, Stern, Kashikawa, \& Spinrad}]{ono2012}
Ono, Y., Ouchi, M., Mobasher, B., {et~al.} 2012, The Astrophysical Journal, 744, 83, \dodoi{10.1088/0004-637X/744/2/83}

\bibitem[{{Ouchi} {et~al.}(2020){Ouchi}, {Ono}, \& {Shibuya}}]{ouchi2020}
{Ouchi}, M., {Ono}, Y., \& {Shibuya}, T. 2020, \araa, 58, 617, \dodoi{10.1146/annurev-astro-032620-021859}

\bibitem[{Ouchi {et~al.}(2008)Ouchi, Shimasaku, Akiyama, Simpson, Saito, Ueda, Furusawa, Sekiguchi, Yamada, Kodama, Kashikawa, Okamura, Iye, Takata, Yoshida, \& Yoshida}]{ouchi2008}
Ouchi, M., Shimasaku, K., Akiyama, M., {et~al.} 2008, The Astrophysical Journal Supplement Series, 176, 301, \dodoi{10.1086/527673}

\bibitem[{{Ouchi} {et~al.}(2018){Ouchi}, {Harikane}, {Shibuya}, {Shimasaku}, {Taniguchi}, {Konno}, {Kobayashi}, {Kajisawa}, {Nagao}, {Ono}, {Inoue}, {Umemura}, {Mori}, {Hasegawa}, {Higuchi}, {Komiyama}, {Matsuda}, {Nakajima}, {Saito}, \& {Wang}}]{ouchi2018}
{Ouchi}, M., {Harikane}, Y., {Shibuya}, T., {et~al.} 2018, \pasj, 70, S13, \dodoi{10.1093/pasj/psx074}

\bibitem[{Pentericci {et~al.}(2011)Pentericci, Fontana, Vanzella, Castellano, Grazian, Dijkstra, Boutsia, Cristiani, Dickinson, Giallongo, Giavalisco, Maiolino, Moorwood, Paris, \& Santini}]{pentericci2011}
Pentericci, L., Fontana, A., Vanzella, E., {et~al.} 2011, The Astrophysical Journal, 743, 132, \dodoi{10.1088/0004-637X/743/2/132}

\bibitem[{Pirzkal {et~al.}(2023)Pirzkal, Rothberg, Papovich, Shen, Leung, Bagley, Finkelstein, Lotz, Koekemoer, Hathi, Cheng, Cleri, Y., Yung, Backhaus, Gardner, {P{\'e}rez-Gonz{\'a}lez}, Ferguson, Grogin, Matharu, Ravindranath, Ryan, Berg, Casey, Castellano, Ortiz, Chworowsky, Dickinson, Somerville, Cox, Dav{\'e}, Davis, {Estrada-Carpenter}, Fontana, Fujimoto, Giavalisco, Grazian, Hutchison, Jaskot, Jung, Kartaltepe, Kewley, Kirkpatrick, Kocevski, Larson, Natarajan, Pentericci, Simons, Snyder, Trump, Vanderhoof, \& Wilkins}]{pirzkal2023}
Pirzkal, N., Rothberg, B., Papovich, C., {et~al.} 2023, The {{Next Generation Deep Extragalactic Exploratory Public Near-Infrared Slitless Survey Epoch}} 1 ({{NGDEEP-NISS1}}): {{Extra-Galactic Star-formation}} and {{Active Galactic Nuclei}} at 0.5 {$<$} z {$<$} 3.6,  arXiv.
\newblock \doarXiv{2312.09972}

\bibitem[{{Planck Collaboration} {et~al.}(2020){Planck Collaboration}, Aghanim, Akrami, Ashdown, Aumont, Baccigalupi, Ballardini, Banday, Barreiro, Bartolo, Basak, Battye, Benabed, Bernard, Bersanelli, Bielewicz, Bock, Bond, Borrill, Bouchet, Boulanger, Bucher, Burigana, Butler, Calabrese, Cardoso, Carron, Challinor, Chiang, Chluba, Colombo, Combet, Contreras, Crill, Cuttaia, {de Bernardis}, {de Zotti}, Delabrouille, Delouis, Di~Valentino, Diego, Dor{\'e}, Douspis, Ducout, Dupac, Dusini, Efstathiou, Elsner, En{\ss}lin, Eriksen, Fantaye, Farhang, Fergusson, {Fernandez-Cobos}, Finelli, Forastieri, Frailis, Fraisse, Franceschi, Frolov, Galeotta, Galli, Ganga, {G{\'e}nova-Santos}, Gerbino, Ghosh, {Gonz{\'a}lez-Nuevo}, G{\'o}rski, Gratton, Gruppuso, Gudmundsson, Hamann, Handley, Hansen, Herranz, Hildebrandt, Hivon, Huang, Jaffe, Jones, Karakci, Keih{\"a}nen, Keskitalo, Kiiveri, Kim, Kisner, Knox, Krachmalnicoff, Kunz, {Kurki-Suonio}, Lagache, Lamarre, Lasenby, Lattanzi, Lawrence, Le~Jeune, Lemos, Lesgourgues,
  Levrier, Lewis, Liguori, Lilje, Lilley, Lindholm, {L{\'o}pez-Caniego}, Lubin, Ma, {Mac{\'i}as-P{\'e}rez}, Maggio, Maino, Mandolesi, Mangilli, {Marcos-Caballero}, Maris, Martin, Martinelli, {Mart{\'i}nez-Gonz{\'a}lez}, Matarrese, Mauri, McEwen, Meinhold, Melchiorri, Mennella, Migliaccio, Millea, Mitra, {Miville-Desch{\^e}nes}, Molinari, Montier, Morgante, Moss, Natoli, {N{\o}rgaard-Nielsen}, Pagano, Paoletti, Partridge, Patanchon, Peiris, Perrotta, Pettorino, Piacentini, Polastri, Polenta, Puget, Rachen, Reinecke, Remazeilles, Renzi, Rocha, Rosset, Roudier, {Rubi{\~n}o-Mart{\'i}n}, {Ruiz-Granados}, Salvati, Sandri, Savelainen, Scott, Shellard, Sirignano, Sirri, Spencer, Sunyaev, {Suur-Uski}, Tauber, Tavagnacco, Tenti, Toffolatti, Tomasi, Trombetti, Valenziano, Valiviita, Van~Tent, Vibert, Vielva, Villa, Vittorio, Wandelt, Wehus, White, White, Zacchei, \& Zonca}]{planckcollaboration2020}
{Planck Collaboration}, Aghanim, N., Akrami, Y., {et~al.} 2020, Astronomy and Astrophysics, 641, A6, \dodoi{10.1051/0004-6361/201833910}

\bibitem[{Qin \& Wyithe(2024)}]{qin2024}
Qin, Y., \& Wyithe, J. S.~B. 2024, Reionization Morphology and Intrinsic Velocity Offsets Allow Transmission of {{Lyman-}}\{{\textbackslash}alpha\} Emission from {{JADES-GS-z13-1-LA}},  arXiv.
\newblock \doarXiv{2409.07356}

\bibitem[{{Qin} {et~al.}(2024){Qin}, {Mesinger}, {Prelogovi{\'c}}, {Becker}, {Bischetti}, {Bosman}, {Davies}, {D'Odorico}, {Gaikwad}, {Haehnelt}, {Keating}, {Lai}, {Ryan-Weber}, {Satyavolu}, {Walter}, \& {Zhu}}]{qin_mesinger2024}
{Qin}, Y., {Mesinger}, A., {Prelogovi{\'c}}, D., {et~al.} 2024, arXiv e-prints, arXiv:2412.00799, \dodoi{10.48550/arXiv.2412.00799}

\bibitem[{Robertson(2022)}]{robertson2022a}
Robertson, B.~E. 2022, Annual Review of Astronomy and Astrophysics, 60, 121, \dodoi{10.1146/annurev-astro-120221-044656}

\bibitem[{Runnholm {et~al.}(2023)Runnholm, Hayes, Lin, Melinder, Scarlata, Adamo, Augustin, Bik, Blaizot, Cannon, Cantalupo, Garel, Gronke, Herenz, Leclercq, {\"O}stlin, Peroux, Rasekh, Rutkowski, Verhamme, \& Wisotzki}]{runnholm2023}
Runnholm, A., Hayes, M.~J., Lin, Y.-H., {et~al.} 2023, Monthly Notices of the Royal Astronomical Society, 522, 4275, \dodoi{10.1093/mnras/stad1264}

\bibitem[{Schechter(1976)}]{Schechter1976}
Schechter, P. 1976, The Astrophysical Journal, 203, 297, \dodoi{10.1086/154079}

\bibitem[{Schenker {et~al.}(2014)Schenker, Ellis, Konidaris, \& Stark}]{schenker2014}
Schenker, M.~A., Ellis, R.~S., Konidaris, N.~P., \& Stark, D.~P. 2014, The Astrophysical Journal, 795, 20, \dodoi{10.1088/0004-637X/795/1/20}

\bibitem[{Schenker {et~al.}(2012)Schenker, Stark, Ellis, Robertson, Dunlop, McLure, Kneib, \& Richard}]{schenker2012}
Schenker, M.~A., Stark, D.~P., Ellis, R.~S., {et~al.} 2012, The Astrophysical Journal, 744, 179, \dodoi{10.1088/0004-637X/744/2/179}

\bibitem[{{Scoville} {et~al.}(2007){Scoville}, {Abraham}, {Aussel}, {Barnes}, {Benson}, {Blain}, {Calzetti}, {Comastri}, {Capak}, {Carilli}, {Carlstrom}, {Carollo}, {Colbert}, {Daddi}, {Ellis}, {Elvis}, {Ewald}, {Fall}, {Franceschini}, {Giavalisco}, {Green}, {Griffiths}, {Guzzo}, {Hasinger}, {Impey}, {Kneib}, {Koda}, {Koekemoer}, {Lefevre}, {Lilly}, {Liu}, {McCracken}, {Massey}, {Mellier}, {Miyazaki}, {Mobasher}, {Mould}, {Norman}, {Refregier}, {Renzini}, {Rhodes}, {Rich}, {Sanders}, {Schiminovich}, {Schinnerer}, {Scodeggio}, {Sheth}, {Shopbell}, {Taniguchi}, {Tyson}, {Urry}, {Van Waerbeke}, {Vettolani}, {White}, \& {Yan}}]{scoville2007}
{Scoville}, N., {Abraham}, R.~G., {Aussel}, H., {et~al.} 2007, \apjs, 172, 38, \dodoi{10.1086/516580}

\bibitem[{{Sharma} {et~al.}(2016){Sharma}, {Theuns}, {Frenk}, {Bower}, {Crain}, {Schaller}, \& {Schaye}}]{sharma2016}
{Sharma}, M., {Theuns}, T., {Frenk}, C., {et~al.} 2016, \mnras, 458, L94, \dodoi{10.1093/mnrasl/slw021}

\bibitem[{{Sobral} \& {Matthee}(2019)}]{sobral2019}
{Sobral}, D., \& {Matthee}, J. 2019, \aap, 623, A157, \dodoi{10.1051/0004-6361/201833075}

\bibitem[{Stark {et~al.}(2010)Stark, Ellis, Chiu, Ouchi, \& Bunker}]{stark2010}
Stark, D.~P., Ellis, R.~S., Chiu, K., Ouchi, M., \& Bunker, A. 2010, Monthly Notices of the Royal Astronomical Society, 408, 1628, \dodoi{10.1111/j.1365-2966.2010.17227.x}

\bibitem[{Stark {et~al.}(2011)Stark, Ellis, \& Ouchi}]{stark2011}
Stark, D.~P., Ellis, R.~S., \& Ouchi, M. 2011, The Astrophysical Journal, 728, L2, \dodoi{10.1088/2041-8205/728/1/L2}

\bibitem[{Tang {et~al.}(2024)Tang, Stark, Topping, Mason, \& Ellis}]{tang2024a}
Tang, M., Stark, D.~P., Topping, M.~W., Mason, C., \& Ellis, R.~S. 2024, {{JWST}}/{{NIRSpec Observations}} of {{Ly}}\${\textbackslash}alpha\$ {{Emission}} in {{Star Forming Galaxies}} at \$6.5{\textbackslash}lesssim Z{\textbackslash}lesssim13\$, \dodoi{10.48550/arXiv.2408.01507}

\bibitem[{{Thai} {et~al.}(2023){Thai}, {Tuan-Anh}, {Pello}, {Goovaerts}, {Richard}, {Claeyssens}, {Mahler}, {Lagattuta}, {de la Vieuville}, {Salvador-Sol{\'e}}, {Garel}, {Bauer}, {Jeanneau}, {Cl{\'e}ment}, \& {Matthee}}]{thai2023}
{Thai}, T.~T., {Tuan-Anh}, P., {Pello}, R., {et~al.} 2023, \aap, 678, A139, \dodoi{10.1051/0004-6361/202346716}

\bibitem[{Tilvi {et~al.}(2020)Tilvi, Malhotra, Rhoads, Coughlin, Zheng, Finkelstein, Veilleux, Mobasher, Wang, Probst, Swaters, Hibon, Joshi, Zabl, Jiang, Pharo, \& Yang}]{tilvi2020}
Tilvi, V., Malhotra, S., Rhoads, J.~E., {et~al.} 2020, The Astrophysical Journal, 891, L10, \dodoi{10.3847/2041-8213/ab75ec}

\bibitem[{{Trenti} \& {Stiavelli}(2008)}]{trentistiavelli2008}
{Trenti}, M., \& {Stiavelli}, M. 2008, \apj, 676, 767, \dodoi{10.1086/528674}

\bibitem[{{Treu} {et~al.}(2022){Treu}, {Roberts-Borsani}, {Bradac}, {Brammer}, {Fontana}, {Henry}, {Mason}, {Morishita}, {Pentericci}, {Wang}, {Acebron}, {Bagley}, {Bergamini}, {Belfiori}, {Bonchi}, {Boyett}, {Boutsia}, {Calabr{\'o}}, {Caminha}, {Castellano}, {Dressler}, {Glazebrook}, {Grillo}, {Jacobs}, {Jones}, {Kelly}, {Leethochawalit}, {Malkan}, {Marchesini}, {Mascia}, {Mercurio}, {Merlin}, {Nanayakkara}, {Nonino}, {Paris}, {Poggianti}, {Rosati}, {Santini}, {Scarlata}, {Shipley}, {Strait}, {Trenti}, {Tubthong}, {Vanzella}, {Vulcani}, \& {Yang}}]{Treu2022ApJ}
{Treu}, T., {Roberts-Borsani}, G., {Bradac}, M., {et~al.} 2022, \apj, 935, 110, \dodoi{10.3847/1538-4357/ac8158}

\bibitem[{{Umeda} {et~al.}(2024{\natexlab{a}}){Umeda}, {Ouchi}, {Nakajima}, {Harikane}, {Ono}, {Xu}, {Isobe}, \& {Zhang}}]{umeda2024}
{Umeda}, H., {Ouchi}, M., {Nakajima}, K., {et~al.} 2024{\natexlab{a}}, \apj, 971, 124, \dodoi{10.3847/1538-4357/ad554e}

\bibitem[{{Umeda} {et~al.}(2024{\natexlab{b}}){Umeda}, {Ouchi}, {Kikuta}, {Harikane}, {Ono}, {Shibuya}, {Inoue}, {Shimasaku}, {Liang}, {Matsumoto}, {Saito}, {Kusakabe}, {Kageura}, \& {Nakane}}]{umeda2024b}
{Umeda}, H., {Ouchi}, M., {Kikuta}, S., {et~al.} 2024{\natexlab{b}}, arXiv e-prints, arXiv:2411.15495, \dodoi{10.48550/arXiv.2411.15495}

\bibitem[{Willis {et~al.}(2008)Willis, Courbin, Kneib, \& Minniti}]{willis2008}
Willis, J.~P., Courbin, F., Kneib, J.~P., \& Minniti, D. 2008, Monthly Notices of the Royal Astronomical Society, 384, 1039, \dodoi{10.1111/j.1365-2966.2007.12404.x}

\bibitem[{Wisotzki {et~al.}(2016)Wisotzki, Bacon, Blaizot, Brinchmann, Herenz, Schaye, Bouch{\'e}, Cantalupo, Contini, Carollo, Caruana, Courbot, Emsellem, Kamann, Kerutt, Leclercq, Lilly, Patr{\'i}cio, Sandin, Steinmetz, Straka, Urrutia, Verhamme, Weilbacher, \& Wendt}]{wisotzki2016}
Wisotzki, L., Bacon, R., Blaizot, J., {et~al.} 2016, Astronomy and Astrophysics, 587, A98, \dodoi{10.1051/0004-6361/201527384}

\bibitem[{Witstok {et~al.}(2024{\natexlab{a}})Witstok, Smit, Saxena, Jones, Helton, Sun, Maiolino, Kumari, Stark, Bunker, Arribas, Baker, Bhatawdekar, Boyett, Cameron, Carniani, Charlot, Chevallard, Curti, {Curtis-Lake}, Eisenstein, Endsley, Hainline, Ji, Johnson, Looser, Nelson, Perna, Rix, Robertson, Sandles, Scholtz, Simmonds, Tacchella, {\"U}bler, Williams, Willmer, \& Willott}]{witstok2024a}
Witstok, J., Smit, R., Saxena, A., {et~al.} 2024{\natexlab{a}}, Astronomy \& Astrophysics, 682, A40, \dodoi{10.1051/0004-6361/202347176}

\bibitem[{Witstok {et~al.}(2024{\natexlab{b}})Witstok, Jakobsen, Maiolino, Helton, Johnson, Robertson, Tacchella, Cameron, Smit, Bunker, Saxena, Sun, Arribas, Baker, Bhatawdekar, Boyett, Cargile, Carniani, Charlot, Chevallard, Curti, {Curtis-Lake}, D'Eugenio, Eisenstein, Hainline, Jones, Kumari, Maseda, {P{\'e}rez-Gonz{\'a}lez}, Rinaldi, Scholtz, {\"U}bler, Williams, Willmer, Willott, \& Zhu}]{witstok2024b}
Witstok, J., Jakobsen, P., Maiolino, R., {et~al.} 2024{\natexlab{b}}, Witnessing the Onset of {{Reionisation}} via {{Lyman-}}\${\textbackslash}alpha\$ Emission at Redshift 13,  arXiv.
\newblock \doarXiv{2408.16608}

\bibitem[{{Witstok} {et~al.}(2025){Witstok}, {Maiolino}, {Smit}, {Jones}, {Bunker}, {Helton}, {Johnson}, {Tacchella}, {Saxena}, {Arribas}, {Bhatawdekar}, {Boyett}, {Cameron}, {Cargile}, {Carniani}, {Charlot}, {Chevallard}, {Curti}, {Curtis-Lake}, {D'Eugenio}, {Eisenstein}, {Hainline}, {Hausen}, {Kumari}, {Laseter}, {Maseda}, {Rieke}, {Robertson}, {Scholtz}, {Shivaei}, {Williams}, {Willmer}, \& {Willott}}]{witstok2025}
{Witstok}, J., {Maiolino}, R., {Smit}, R., {et~al.} 2025, \mnras, 536, 27, \dodoi{10.1093/mnras/stae2535}

\bibitem[{Wold {et~al.}(2022)Wold, Malhotra, Rhoads, Wang, Hu, Perez, Zheng, Khostovan, Walker, Barrientos, {Gonz{\'a}lez-L{\'o}pez}, Harish, Infante, Jiang, Pharo, {Moya-Sierralta}, Bauer, Galaz, Valdes, \& Yang}]{wold2022}
Wold, I. G.~B., Malhotra, S., Rhoads, J., {et~al.} 2022, The Astrophysical Journal, 927, 36, \dodoi{10.3847/1538-4357/ac4997}

\bibitem[{{Zhu} {et~al.}(2024){Zhu}, {Becker}, {Bosman}, {Cain}, {Keating}, {Nasir}, {D'Odorico}, {Ba{\~n}ados}, {Bian}, {Bischetti}, {Bolton}, {Chen}, {D'Aloisio}, {Davies}, {Davies}, {Eilers}, {Fan}, {Gaikwad}, {Greig}, {Haehnelt}, {Kulkarni}, {Lai}, {Puchwein}, {Qin}, {Ryan-Weber}, {Satyavolu}, {Spina}, {Walter}, {Wang}, {Wolfson}, \& {Yang}}]{zhu2024}
{Zhu}, Y., {Becker}, G.~D., {Bosman}, S. E.~I., {et~al.} 2024, \mnras, 533, L49, \dodoi{10.1093/mnrasl/slae061}

\end{thebibliography}
\end{document}